\documentclass[12pt,a4]{article}
\usepackage{latexsym}
\usepackage{mathrsfs}
\usepackage{amssymb}
\usepackage{amsmath}
\usepackage{graphicx}
\usepackage{braket}
\usepackage{enumerate}
\usepackage[usenames,dvipsnames]{xcolor}
\usepackage{todonotes}

\newcommand{\A}{\mathbb{A}}
\newcommand{\C}{\mathbb{C}}
\newcommand{\HH}{\mathbb{H}}

\newcommand{\R}{\mathbb{R}}

\newcommand{\scri}{\mathscr{I}}

\newcommand{\T}{\mathbb{T}}

\newcommand{\p}{\partial}
\newcommand{\dbar}{\bar\partial}
\newcommand{\e}{\mathrm{e}}

\newcommand{\cA}{\mathcal{A}}

\newcommand{\cM}{\mathcal{M}}
\newcommand{\cN}{\mathcal{N}}
\newcommand{\cO}{\mathcal{O}}
\newcommand{\cV}{\mathcal{V}}

\newcommand{\SL}{\mathrm{SL}}

\newcommand{\vol}{\mathrm{Vol}}

\newcommand{\SU}{\, \mathrm{SU}}
\newcommand{\GL}{\mathrm{GL}}

\newcommand{\rd}{\, \mathrm{d}}

\newcommand{\be}{\begin{equation}\label}
\newcommand{\ee}{\end{equation}}
\newcommand{\bea}{\begin{eqnarray}\label}
\newcommand{\eea}{\end{eqnarray}}

\newcommand{\la}{\langle}
\newcommand{\ra}{\rangle}

\newcommand{\bd}{\bar{\delta}}
\newcommand{\wt}{\widetilde}

\begin{document}

\title{Ambitwistor strings at null infinity and subleading soft limits}
\author{Yvonne Geyer, Arthur E. Lipstein, Lionel Mason \vspace{7pt}\\ \normalsize \textit{
Mathematical Institute, Andrew Wiles Building,}\\\normalsize\textit{Woodstock Road, Oxford, OX2 6CG, United Kingdom}}

\maketitle

\begin{abstract}
The relationships between extended BMS symmetries at null infinity and
Weinberg's soft theorems for gravitons and photons together with their subleading generalizations are developed using ambitwistor string theory.   
Ambitwistor space is the phase space of
complex null geodesics in complexified space-time.  We show how it
can be canonically identified with the cotangent bundle of null infinity.
BMS symmetries of null infinity lift to give a hamiltonian action on ambitwistor space, both in general dimension and in its twistorial 4-dimensional representation.  
General vertex operators arise from hamiltonians generating
diffeomorphisms of ambitwistor space that determine the scattering from past to future null
infinity. When a momentum eigenstate goes soft, the diffeomorphism defined by its leading and
its subleading part are extended BMS generators realized in the world
sheet conformal field theory of the ambitwistor string.  More
generally, this gives explicit perturbative correspondence
between the scattering of null geodesics and that of the gravitational
field via ambitwistor string theory.
\end{abstract}

\newpage
\tableofcontents

\section{Introduction}
It has long been understood that infrared behaviour in gravity is related to supertranslation ambiguities in the choices of coordinates at null infinity \cite{Ashtekar:1981sf}. 
In a recent series of papers
\cite{Strominger:2013lka,Strominger:2013jfa,He:2014laa,Cachazo:2014fwa},
Strominger and coworkers have proposed a new way of understanding the
Weinberg soft limit theorems in terms of the action of the BMS group 
 at null infinity and used the approach to suggest new
theorems for the subleading terms in the soft limit.  In an intriguing
recent paper \cite{Adamo:2014yya}, Adamo, Casali and Skinner proposed
a string model at null infinity for four dimensions to provide an
explanation for these ideas.  They also suggested a link
with ambitwistor strings that extends their ideas to arbitrary
dimension, particularly in view of the recent proof of the subleading
soft limit results in \cite{Schwab:2014xua} using the formulae of
Cachazo, He and Yuan \cite{Cachazo:2013gna, Cachazo:2013hca,
  Cachazo:2013iea} as these formulae arise from ambitwistor string theory in
arbitrary dimensions \cite{Mason:2013sva}. 

 The
purpose of this paper is to explain the relationship of ambitwistor
strings to null infinity and to prove the  soft factor
theorems from the perspective of the conformal field theory that
arises from representing ambitwistor string theories at null infinity.  We
also make clear the relationship between diffeomorphisms of null
infinity and the vertex operators in the theory, and how a
soft momentum eigenstate 
becomes an extended BMS generator at leading and subleading order.  We will also see that these ideas are realized straightforwardly in the more twistorial four dimensional ambitwistor strings \cite{Geyer:2014fka, Geyer:2014} and that indeed this model is closely related  to that of \cite{Adamo:2014yya}.

Several decades ago, Weinberg showed that photon and graviton
amplitudes behave in a universal way when one of the external
particles with momentum $s$ becomes soft \cite{Weinberg:1965nx}: 
\[
\mathcal{A}_{n+1}\rightarrow\sum_{a=1}^{n}\frac{\epsilon_{a}\cdot
  k_{a}}{s\cdot
  k_{a}}\mathcal{A}_{n},\qquad \mathcal{M}_{n+1}\rightarrow\sum_{a=1}^{n}\frac{\epsilon_{\mu\nu}k_{a}^{\mu}k_{a}^{\nu}}{s\cdot
  k_{a}}\mathcal{M}_{n}\, , \qquad \mbox{as } s\rightarrow 0 
\]
Strominger and collaborators have shown that these formulae
are a consequence of the asymptotic symmetries of Minkowski
space, known as the BMS group
\cite{Strominger:2013lka,Strominger:2013jfa,He:2014laa}. The BMS group
are diffeomorphisms of null infinity $\scri$ that preserve its weak
and strong conformal structure \cite{Penrose:1986ca} in four dimensions.
The null infinity of $d$-dimensional Minkowski space $\scri\cong
\mathbb{R}\times S^{d-2}$ is a light cone, the product of a conformal
$d-2$-sphere with the null generators $\R$. The BMS group consists of
global conformal transformations of the $d-2$-sphere and so-called
supertranslations up the generators
\cite{Bondi:1962px,Sachs:1962wk}. Strominger et.\ al.\ argue that the
soft limit theorem of Weinberg arises from the Ward identity following
from supertranslation invariance, but taking only a diagonal subgroup of the product of the groups obtained at past null infinity $\scri^-$ with that at $\scri^+$.  This diagonal subgroup is obtained by a requiring the real part of the second derivative of the shear from $\scri^-$ to be equal at space-like infinity to that from $\scri^+$.   Furthermore, the BMS group can be
extended by so-called superrotations, which correspond to extending
the global conformal symmetry of the 2-sphere in $d=4$ to a local conformal symmetry \cite{Barnich:2009se,Barnich:2011ct,Barnich:2011mi} in 4-dimensions (and more general diffeomorphisms in higher dimensions). 

Recently, Cachazo and Strominger discovered  subsubleading terms in the soft limit of tree-level graviton amplitudes in four dimensions,
\begin{equation}
 \cM_{n+1}=\left(S^{(0)}+S^{(1)}+S^{(2)}\right)\cM_n+\cO(s^2)\,,
\end{equation}
where the contributions to leading and subleading order are given by
\begin{align*}
 S^{(0)}=\sum_{a=1}^{n}\frac{(\epsilon\cdot k_a)^2}{s\cdot k_{a}}, \qquad S^{(1)}=\frac{\epsilon_{\mu\nu}k_{a}^{\mu}s_{\lambda}J_{a}^{\lambda\nu}}{s\cdot k_{a}}, \qquad
 S^{(2)}=\frac{\epsilon_{\mu\nu}(s_{\lambda}J_{a}^{\lambda\mu})(s_{\rho}J_{a}^{\rho\nu})}{s\cdot k_{a}}\,.
\end{align*}
Moreover, they conjectured that the subleading term follows from superrotation symmetry and should therefore also be universal \cite{Cachazo:2014fwa}.  A similar subleading factor was found by Casali for tree-level Yang-Mills amplitudes in four dimensions \cite{Casali:2014xpa};
\begin{equation}
 \cA_{n+1}=\left(S^{(0)}+S^{(1)}\right)\cA_n+\cO(s)\,,
\end{equation}
where $S^{(0)}$ denotes the Weinberg soft limit, and $S^{(1)}$ is the subleading contribution,
\begin{align*}
 S^{(0)}=\sum_{a\text{ adj. }s}\frac{\epsilon\cdot k_{a}}{s\cdot k_{a}}, \qquad S^{(1)}=\sum_{a\text{ adj. }s} \frac{\epsilon_{\mu}s_{\nu}J_{a}^{\mu\nu}}{s\cdot k_{a}}\,.
\end{align*}
Schwab and Volovich subsequently proved these subleading soft limit formulae for tree-level Yang-Mills and gravity amplitudes in any spacetime dimension using the formulae of Cachazo, He, and Yuan (CHY) \cite{Schwab:2014xua}. The CHY formulae describe tree-level scattering of massless particles in arbitrary dimensions and are based on the so-called scattering equations \cite{Cachazo:2013iaa,Cachazo:2013gna,Cachazo:2013hca,Cachazo:2013iea}. Loop corrections to the subleading soft limits were subsequently studied using dimensional regularization in \cite{Bern:2014oka,He:2014bga,Cachazo:2014dia}. 

Adamo, Casali, and Skinner subsequently proposed a 2d CFT living on the complexification of $\scri$, whose correlation functions give the tree-level amplitudes of $\mathcal{N}=8$ supergravity \cite{Adamo:2014yya}. They also defined charges whose insertions into correlators give rise to leading and subleading soft limits discovered by Strominger and Cachazo, and therefore correspond to generators of supertranslations and superrotations. They tentatively suggested some relations with ambitwistor string theories.

Ambitwistor string theories provide a powerful point of view on the tree-level gauge and gravity amplitudes. They naturally encode the scattering equations and give rise to the CHY formulae in arbitrary dimensions \cite{Mason:2013sva} being critical in ten. The geometric framework was used to develop 4d ambitwistor strings \cite{Geyer:2014fka, Geyer:2014} now in a twistorial representation rather than the original vector representation so that supersymmetry can be naturally explicitly incorporated.  They give rise to new twistor string formulae for tree-level Yang-Mills and gravity amplitudes with any amount of supersymmetry that are much simpler than previous formulae. We will see that the 2d CFT of Adamo, Casali \& Skinner is closely related to this four-dimensional ambitwistor string.

Here we explain how ambitwistor space can be identified with the cotangent bundle of null infinity in such a way that the BMS
generators and their generalizations, indeed arbitrary symplectic diffeomorphisms of $T^*\scri$, act canonically. This leads directly to an action on the worldsheet theory of the ambitwistor string.   We will see that a general integrated vertex operator corresponding to a  graviton insertion can be expressed as  the implementation of such a diffeomorphism in the worldsheet theory.  
Such a vertex operator for a momentum eigenstate can
be straightforwardly expanded in powers of the soft momentum. In the soft limit for gravity we obtain worldsheet generators of supertranslations on $\scri$ to leading order and superrotations at subleading order.  In the expansion of the vertex operator we obtain operators that generate an infinite series of new higher-order soft terms corresponding to more general diffeomorphisms of $T^*\scri$.

The analagous story for Yang-Mills is that vertex operators at null infinity correspond
to certain gauge transformations at  $T^*\scri$.  In its soft expansion, we obtain  gauge transformations
analagous to supertranslations at leading order and superrotations for the subleading terms.

In appendices we give detailed calculations of the correlators of the soft graviton and Yang-Mills vertex operators with a general set of vertex operators to give proofs of the soft limits for the leading, subleading and some sub-subleading terms in the form required of the Ward identity arguments of Strominger et.\ al..

\section{Ambitwistor strings at null infinity}
\subsection{Background geometry}
Ambitwistor space $\A$ is the complexification of the phase  space of
complex null geodesics with scale in a space-time.  As such, they can
be represented by their directions and their intersection with any
Cauchy surface.  The symplectic potential $\Theta$ and symplectic form
$\rd\Theta$ on $\A$ arise from identifying $\A$ with the cotangent
bundle of the complexification of that Cauchy  hypersurface.  In an
asymptotically simple space-time,  they can therefore be represented
with respect to the complexification of null infinity, which we will
denote  $\scri$, and so $\A=T^*\scri$; and at this point $\scri$ can be
the complexification of either future or past null infinity, $\scri^+$
or $\scri^-$.

\begin{figure}[htbp] 
\centering
       \includegraphics[width=3.2in]{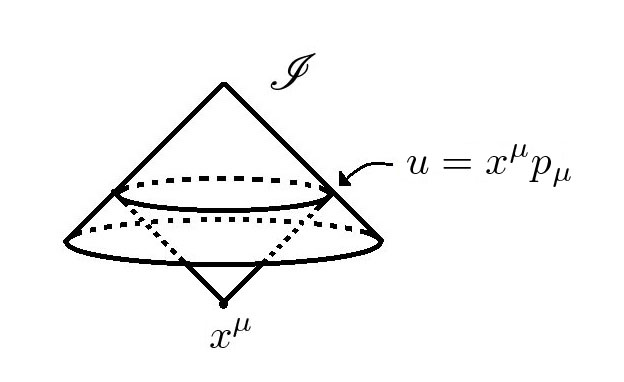}  
    \caption{Diagram of null infinity, $\scri$.}
    \label{scri}
    \end{figure}

Null infinity can be represented as a light cone, although it is normal to invert the parameter up the generators to give a parameter $u$ for which the vertex is at $u=\infty$.  In order to make the symmetries manifest, we use homogeneous coordinates $p_\mu$ with $p^2=0$ for the complexified sphere of generators of $\scri$, and a coordinate $u$ of weight one also, so that $(u, p_\mu)\sim (\alpha u, \alpha p_\mu)$ for $\alpha\neq 0$.  As depicted in Figure \ref{scri}, a null geodesic through a point $x^\mu$ with null tangent vector $P_\mu$ reaches $\scri$ at the point with coordinates 
$$
(u,p_\mu)=w (x^\mu P_\mu,P_\nu)\, ,
$$
where $w$ encodes the scale of $P$.  The notation is intended to be
suggestive of the fact that $u$ is canonically conjugate to frequency
here denoted $w$.
Since  $\A=T^*\scri$, we can give homogeneous coordinates $(u,p_\mu, w,q^\mu)$ with $(w, q^\mu)$ of weight $0$ and $(u,p_\mu)$ weight one to yield the 1-form 
\begin{equation}
\Theta= w \rd u- q^\mu \rd p_\mu\, 
\end{equation}
and this defines the symplectic potential on $\A$.
As $\Theta $ must be orthogonal to the Euler vector field $\Upsilon= u\p_u +p_\mu\p_{p_\mu}$  we have the constraint
\be{constraint}
wu-q\cdot p=0\, ,
\ee
which is the hamiltonian for $\Upsilon$.

In \cite{Mason:2013sva}, the coordinate description of $\A$ was given as the symplectic quotient of the cotangent bundle of space-time, i.e.\  as $(x^\mu, P_\mu)$ with $P^2=0$, quotiented by $P\cdot \p_{x}$.  In including the scale, $\A$ is a symplectic manifold with symplectic form $\rd \Theta $ where $\Theta=P\cdot \rd x$.
The null geodesic through $x^\mu$ with null cotangent vector $P_\mu$ has coordinates determined by $u=p\cdot x$ and 
$$
\Theta =P_\mu \rd x^\mu =w\rd u-q^\mu \rd p_\mu 
$$
so that 
$$
q^\mu=w x^\mu \mbox{ modulo } p^\mu\, ,  \quad \mbox{ and } \quad p=w P\, .
$$
On reducing by the constraint \eqref{constraint}, we can identify the scalings of $p$ with those of the momentum by scaling $w$ to 1.  For scaled null geodesics, we can therefore simply incorporate the scale of $p$.
The above  new coordinates for $\A$  in effect just add the two new coordinates $(u,w)$ to the original description, which can be eliminated by performing the reduction by the constraint \eqref{constraint} and making the gauge choice $w=1$  in the quotient.  However, they serve to reveal its relationship with $\scri$ and to express $\A$ as the symplectic quotient of $(u,p,w,q)$ space by the constraints $p^2=0$ and $uw-q\cdot p=0$.

For the RNS ambitwistor string models, we augment the coordinates above to include either $d$ or $2d$ fermionic coordinates respectively in the heterotic case and the type II case.  These are given by coordinates $\Psi^\mu_r$, $r=1,2$ in the type $II$ case (and $r=1$ in the heterotic case), subject to the constraint(s)  $wp\cdot \Psi_r=0$.  The symplectic potential is then augmented to
\be{susy-pot}
\Theta=w\rd u - q\cdot \rd p + \Psi\cdot \rd \Psi\, ,
\ee
and symplectic form $\rd\Theta$.
\subsection{BMS symmetries and their generalizations}
All diffeomorphisms of a manifold have a hamiltonian lift to the cotangent bundle with Hamiltonian given by the contraction of the generating vector field with the symplectic potential  $\Theta$.

Poincar\'e motions in particular act as diffeomorphisms of $\scri$. 
Translations act by $\delta x^\mu=a^\mu$ give $\delta p=0$, $\delta u=a\cdot p$ and $\delta q^\mu=w a^\mu$ and have Hamiltonian 
\begin{equation}
H_a=wa\cdot p\, .
\end{equation}
 Supertranslations generalize these to $\delta u= f(p)$ where $f$ is now an arbitrary function of weight 1 in $p$ (i.e., a section of $\cO(1)$) but no longer necessarily linear (and generally with singularities).
These motions are all symplectic with hamiltonian
\begin{equation}
H_f=w f(p)\, .
\end{equation}

Lorentz transformations  act by $\delta p_\mu=r_{\mu}{}^\nu p^\nu$, $\delta q^\mu=-r_\nu{}^\mu q^\nu$ ( similarly for  $x^\mu$ and $\Psi_r^\mu$) with $r_{\mu\nu}=r_{[\mu\nu]}$.  
This action has a natural lift to the total space of the line bundle $\cO(1)$ of homegeneity degree 1 functions in which $u$ takes its values.
The hamiltonian for this action is 
\begin{equation}
H_r=  (q^{[\mu} p^{\nu]} +w\sum_r \Psi_r^\mu\Psi_r^\nu) r_{\mu\nu}\, .
\end{equation}
We can define the angular momentum to be
\begin{equation}\label{angmom}
J^{\mu\nu}=(q^{[\mu} p^{\nu]} +w\sum_r \Psi_r^\mu\Psi_r^\nu)\, .
\end{equation}
It is the sum of an orbital part and an intrinsic spin part and commutes with the constraints $p^2=0$ and $wp\cdot \Psi_r=0$.

Superrotations will be defined here by generalizing $r_{\mu\nu}$ to functions that have non-trivial dependence on $p$ (but still of weight zero),
\be{superr}
H_r=  J^{\mu\nu} r_{\mu\nu}(p)\, . 
\ee
They also preserve the constraints $p^2=p\cdot \Psi=0$ on the constraint surface.
In general dimension conformal motions are finite dimensional even locally, but in four, if not constrained to be global on the Riemann sphere, become infinite dimensional and provide a nontrivial restriction on the general diffeomorphisms we have allowed above.

\subsection{The string model}
As in the original ambitwistor string, the action is determined by the symplectic potential $\Theta$.  This gives the worldsheet action on a Riemann surface $\Sigma$ in the new coordinates as
\begin{multline}
S= \frac1{2\pi}\int_\Sigma w\dbar u -q^\mu \dbar p_\mu + \Psi_r\cdot \dbar \Psi _r +\\ 
+ eT +  \tilde e p^2 + \chi_r w p\cdot\Psi_r + a(uw-q\cdot p)\, .
\end{multline}
Here we have taken $(u,p_\mu)$ to take values in $\kappa=\Omega^{1,0}$ and  $\Psi_r$  in a worldsheet spin bundle $\kappa^{1/2}$ and $(w,q^\mu)$ are worldsheet scalars. $\tilde e\in \Omega^{0,1}\otimes T$, $\xi_r \in \Omega^{0,1}\otimes T^{1/2}$ and $a\in \Omega^{0,1}$ are gauge fields imposing and gauging the various constraints. The term $eT$ with $T=(w\p u -q^\mu \p p_\mu + \Psi_r\cdot \p \Psi _r) $  allows for an arbitrary choice of complex structure parametrized by $e$.  

These gaugings are fixed by setting $e=\tilde e=\chi_r=a=0$ but lead to
respective ghost systems $(b,c)$ and $(\tilde b,\tilde c)$ fermionic, $(\beta_r
,\gamma_r)$ bosonic and $(r,s)$ fermionic.  We are left with the
BRST operator
\be{Q}
Q_{BRST}=\frac1{2\pi i}\oint c T+ \tilde c \frac {p^2}2 + \gamma_r w\Psi_r\cdot P + s
(uw-q\cdot p)\, .
\ee

This is sufficiently close to the original ambitwistor string that we
can simply adapt the Yang-Mlls vertex operators in the heterotic model (with $r=1$ only) and the gravitational vertex operators in the type II case with $r=1,2$. With momentum vector $k^\mu$ and polarization vectors
$\epsilon_{r\mu}$ to give 
\begin{align}\label{Vertex}
U&= \e^{ik\cdot q/w}\prod_{r=1}^2\delta(\gamma_r) \Psi_r\cdot \epsilon_r\, ,\\
\cV&=\int_\Sigma \bar\delta(k\cdot p) \, w \,\e^{ik\cdot q/w}
\prod_{r=1}^2\epsilon_{r\mu}(p^\mu +i\Psi_r^\mu \Psi_r\cdot k)\, ,
\end{align}
for gravity. For Yang-Mills we have
\begin{align}\label{VertexYM}
U^{ym}&= \e^{ik\cdot q/w}\delta(\gamma) \Psi\cdot \epsilon j\cdot t\, , \\
\cV^{ym}&=\int_\Sigma \bar\delta(k\cdot p) w  \e^{ik\cdot q/w}
\epsilon_{\mu}(p^\mu +i\Psi^\mu \Psi\cdot k) j\cdot t\, .
\end{align}
where $j$ is a current algebra on the worldsheet associated to the gauge group and $t$ a lie algebra element.  As described in \cite{Mason:2013sva}, $\cV$ are the integrated vertex operators, and $U$ are unintegrated
with respect to both the zero modes of $\gamma_r$ and
$\tilde c$.  When $\Sigma $ is a Riemann sphere, we need two
insertions of $c\tilde c U$ to fix the two pairs of $\gamma_r$
zero-modes and and a third insertion of $c\tilde c $ multuplied by an unintegrated version of $\cV$ (without its $\bar\delta(k\cdot p)$) to fix the third
of the $c$ and $\tilde c$ zero-modes.

The new feature here is the gauge field $a$ whose ghost $s$ has a zero
mode that must also be fixed.  This can be associated also with a
$1/\mathrm{Vol GL}(1)$ factor from the scalings and we will treat this
as the requirement that the scale of $w$ be fixed to be 1.  This can
be done before correlators are taken because the vertex operators do
not depend on $u$.  At this
point it is easily seen that the amplitude computations directly reduce to
those of \cite{Mason:2013sva} to yield the CHY formulae.
A key feature of this derivation is that, in the evaluation of the correlation functions, the exponentials in the vertex operators are taken into the off-shell action leading to the following expression for $p$ 
\be{p}
p(\sigma)=\sum_i \frac{k_i}{\sigma-\sigma_i}\, ,
\ee
on the moduli space.

\subsection{Symmetries, vertex operators and diffeomorphisms} \label{symmetry}
Because the action of the worldsheet model is based on the symplectic potential, the singular parts of OPE of operators in the ambitwistor string theory precisely arise from the Poisson structure, so that for example
\begin{equation}
p_\mu(\sigma') q^\nu(\sigma) \sim \frac {\delta^\nu_\mu \rd\sigma}{\sigma-\sigma'} + \ldots \, , \qquad ,\Psi^{\mu}(z)\Psi_{\nu}(w)=\frac{\delta_{\nu}^{\mu}}{z-w}+...\label{eq:ope}
\end{equation}
where the ellipses denote finite terms.
The hamiltonians must all have weight one in $p$ (or weight one in $\Psi_r$) to preserve the symplectic potential and so on the worldsheet they take vaules in $\Omega^{1,0}_\Sigma$.   We can therefore directly use the hamiltonian $h$ that generates a symplectic diffeomorphism of  $\A$  to define an operator
$$
Q_h=\frac1{2\pi i}\oint h
$$ 
that induces the action of the symplectic diffeomporphism in the ambitwistor string model, i.e.\ for translations we have
$$
Q_{p_\mu} q^{\nu}=\frac1{2\pi i}\oint  \frac {\delta^\nu_\mu}{\sigma-\sigma'} + \ldots = \delta^\nu_\mu\, .
$$
Clearly the same logic will apply to more general BMS transformations and indeed more general diffeomorphisms of $\scri$ as these all have a symplectic lift to $\A=T^*\scri$.

In fact all vertex operators can be related to such motions.  This is most easily stated for gravity where we can rewrite the integrated vertex operator as 
\begin{eqnarray}
\cV&=&\int_\Sigma \bar\delta(k\cdot p)\, w\,  \e^{ik\cdot q/w}
\prod_{r=1}^2\epsilon_{r\mu}(p^\mu +i\Psi_r^\mu \Psi_r\cdot k)\, ,\nonumber \\ 
&=& \frac1{2\pi i}\oint \frac{\e^{ik\cdot q/w}}{k\cdot p} \, w\, 
\prod_{r=1}^2\epsilon_{r\mu}(p^\mu +i\Psi_r^\mu \Psi_r\cdot k) \, ,\label{gen-diffeo}
\end{eqnarray}
where we have used the relation
$$
\bar\delta(k\cdot p)= \frac1{2\pi i}\dbar\frac1{k\cdot p}
$$ 
to reduce the integral over $\Sigma$ to a contour integral around the pole at $p\cdot k=0$. Thus we see that the vertex operator is the generator of the diffeomorphism of $\A$ with hamiltonian given by the integrand of \eqref{gen-diffeo}.  This is to be expected in the ambitwistor construction as the data of the space-time metric is encoded in deformations of the complex structure of ambitwistor space \cite{LeBrun:1983}.  Such deformations can in turn be encoded in a Dolbeault fashion as a global variation of the $\dbar$-operator as in the first line of \eqref{gen-diffeo} or as a \v Cech deformation of of the patching functions for the manifold as determined by the hamiltonian in the second line.

The story for Yang-Mills is very similar except that now we are talking about variations of the $\dbar$-operator on a bundle in the Dolbeault description, or a non-global gauge transformation in the \v Cech description. In particular, we can rewrite the integrated vertex operator as
\begin{equation}
\mathcal{V}^{ym}=\frac{1}{2\pi i}\oint\frac{\e^{ik\cdot q/w}}{k\cdot p} w \;\epsilon_{\mu}(p^{\mu}+i\Psi^{\mu}\Psi\cdot k)\, j
\label{gen-ym}
\end{equation}
where $j$ is the worldsheet current algebra.

\section{From soft limits to BMS}

In this section, we will expand the gravitational vertex operator in \eqref{gen-diffeo} in the soft limit, which corresponds to the momentum of the graviton going to zero, and show that the leading and subleading terms in the expansion correspond to generators of supertranslations and superrotations, respectively.

Denoting the soft momentum as $s$, we can expand the vertex operator as follows:
\begin{eqnarray} 
 \cV_s&=&\frac1{2\pi i}\oint  \frac{w\, \e^{is\cdot q/w}}{s\cdot p} \prod_{r=1}^2 \epsilon_{r\mu}(p^\mu+i\Psi_r^\mu\Psi_r\cdot s) \nonumber \\
&= & \cV_{s}^0+ \cV_{s}^1+\cV_{s}^2+\cV_{s}^3 +\ldots.
\end{eqnarray}
Simplifying to the situation where $\epsilon_1=\epsilon_2$ (which is sufficient for ordinary gravity), the first two terms in the expansion are given by
\begin{align} \label{expgravsoftVO}
\cV_{s}^0&= \frac1{2\pi i}\oint w\frac{ (\epsilon\cdot p)^2}{s\cdot p} \nonumber \\
\cV_{s}^1&= \frac 1{2\pi i}\oint \frac{\epsilon\cdot p}{s\cdot p} \left ({  i\epsilon\cdot p\,   s\cdot q} + iw\sum_{r=1}^2  \epsilon\cdot\Psi_r s\cdot \Psi_r \right)  \nonumber \\
&=\frac 1{2\pi }\oint \frac{\epsilon\cdot p}{s\cdot p} \epsilon^\mu s^\nu\left ({  p_{[\mu}\,   q_{\nu]}}+ w\sum_{r=1}^2  \Psi_{r\mu}  \Psi_{r\nu} \right) \nonumber \\
&=\frac 1{2\pi }\oint \frac{\epsilon\cdot p}{s\cdot p} \epsilon^\mu s^\nu J_{\mu\nu},
\end{align}
where we have used the angular momentum operator defined in \eqref{angmom},
\be{J}
J_{\mu\nu}={ p_{[\mu}q_{\nu]}} +w\sum_{r=1}^2  \Psi_{r\mu}  \Psi_{r\nu},
\ee 
which corresponds to a sum of orbital angular momentum and intrinsic spin. To get to the second line of \eqref{expgravsoftVO}, we note that the extra $s\cdot p$ term in the numerator cancels that in the denominator and so there is no singularity and the contour integral gives zero. Note that the integrands of these terms in the soft expansion correspond precisely to the generators of the hamiltonian lift of the supertranslations and superrotations of null infinity discussed in section \ref{symmetry}. In particular, $\cV_s^{0}$ generates the supertranslation $\delta u= \frac{(\epsilon\cdot p)^2}{s\cdot p}$, and $\cV_s^{1}$ generates the superrotation $r_{\mu\nu}= \epsilon_{[\mu}s_{\nu]}\frac{\epsilon\cdot p}{s\cdot p}$ on $\scri$.

By a similar calculation to that for $\cV_{s}^1$, one finds that
\begin{align}
\cV_{s}^2 &= \frac1{2\pi i} \oint \frac{(\epsilon\cdot p)^2 (s\cdot q)^2 +2w \epsilon\cdot p s\cdot q \sum_r \epsilon\cdot \Psi s\cdot \Psi + w^2\prod_r \epsilon\cdot \Psi_r s\cdot \Psi_r}{2w\;s\cdot p}\, ,\nonumber\\ 
&= \frac1{2\pi i} \oint \frac{( \epsilon^\mu s^\nu J_{\mu\nu})^2}{2w\;s\cdot p}.
\end{align}  
$\cV_{s}^2$ therefore gives a `superrotation squared' on ambitwistor space. Note that $\cV_{s}^2$ does not generate a symmetry of null infinity, since the square of a symmetry generator does not in general correspond to another symmetry generator. Hence, beyond subleading order, terms in the expansion of a soft graviton vertex operator generate diffeomorphisms of ambitwistor space $\A=T^*\scri$, but not diffeomorphisms of $\scri$ itself. 

In appendix \ref{gravd}, we show that correlators of $\mathcal{V}_s^{0}$ and $\mathcal{V}_s^{1}$ give rise to the leading and subleading terms in the soft limit of graviton amplitudes:
\begin{align}
\left\langle \mathcal{V}_{1}...\mathcal{V}_{n} \cV_{s}^0\right\rangle &=\left(\sum_{a=1}^{n}\frac{(\epsilon\cdot k_a)^2}{s\cdot k_{a}}\right)\left\langle \mathcal{V}_{1}...\mathcal{V}_{n}\right\rangle  \label{Ward0}\\
\left\langle \mathcal{V}_{1}...\mathcal{V}_{n} \cV_{s}^1\right\rangle &=\sum_{a=1}^{n}\frac{\epsilon_{\mu\nu}k_{a}^{\mu}s_{\lambda}J_{a}^{\lambda\nu}}{s\cdot k_{a}}\left\langle \mathcal{V}_{1}...\mathcal{V}_{n}\right\rangle  \label{Ward1}
\end{align}
where $\epsilon^{\mu\nu}=\epsilon^{\mu}\epsilon^{\nu}$
and $J_a^{\mu\nu}=k_{a}^{[\mu}\frac{\partial}{\partial k_{a,\nu]}}+\epsilon_{a}^{[\mu} k_{a}^{\nu]}$. These results are consistent with the claims of Strominger et al \cite{Strominger:2013jfa,He:2014laa,Cachazo:2014fwa} that the soft theorems are equivalent to Ward identities associated with the diagonal subgroup of $BMS^+ \otimes BMS^-$, which is conjectured to be a symmetry of the gravitational S-matrix. In showing that the Weinberg soft theorem follows from this Ward identity, a key step in the argument of Strominger et al is that acting with a supertranslation on null infinity leads to the insertion of a soft graviton, and similar considerations should apply for superrotations. This is precisely what we have demonstrated. 

Equations \eqref{Ward0} and \eqref{Ward1} contain all the information encoded in the Ward identities for supertranslations and superrotations. The correlation functions with insertions of $\cV_s^0$ and $\cV_s^1$, yielding the leading and subleading contributions for graviton amplitudes, imply the general Ward identities for arbitrary supertranslations and superrotations with hamiltonians $H_f$ and $H_r$. \\

There is a similar story for Yang-Mills theory. If we expand the gluon vertex operator in \eqref{gen-ym} in powers of the soft momentum $s$, we obtain the series
\begin{eqnarray} 
 \cV^{ym}_s&=&\frac1{2\pi i}\oint  \frac{\e^{is\cdot q/w}}{s\cdot p}  \epsilon_{\mu}(p^\mu+i\Psi^\mu\Psi \cdot s) \, j \nonumber \\
&= & \cV^{ym,0}_{s}+ \cV^{ym,1}_{s}+\cV^{ym,2}_{s}+\cV^{ym,3}_{s} +\ldots
\end{eqnarray}
where
\begin{align} \label{ymsoftser}
\cV^{ym,0}_{s}&= \frac1{2\pi i}\oint \frac{ \epsilon\cdot p}{s\cdot p}\, j \nonumber \\
\cV^{ym,1}_{s}&=\frac 1{2\pi }\oint \frac{\epsilon^\mu s^\nu}{s\cdot p}  J_{\mu\nu} \, j.
\end{align}
Hence, the leading and subleading terms in the expansion of the gluon vertex operator generate an analogue of supertranslations and superrotations for Yang-Mills theory being respectively generators of gauge transformations that are depend only on $p$ or linear in $J_{\mu\nu}$. Unlike  gravity, $\cV^{ym,2}_{s}$ is no longer the square of $J$. 

In appendix \ref{ymd}, we show that correlators of $\mathcal{V}^{ym,0}_{s}$ and $\mathcal{V}^{ym,1}_{s}$ give rise to the leading and subleading terms in the soft limit of gluon amplitudes: 
\begin{align}
\left\langle \mathcal{V}_{1}...\mathcal{V}_{n} \mathcal{V}^{ym,0}_{s} \right\rangle &=\left(\frac{\epsilon\cdot k_{1}}{s\cdot k_{1}}-\frac{\epsilon\cdot k_{n}}{s\cdot k_{n}}\right)\left\langle \mathcal{V}_{1}...\mathcal{V}_{n}\right\rangle \nonumber \\
  \left\langle \mathcal{V}_{1}\dots\mathcal{V}_{n}\mathcal{V}^{ym,1}_{s}\right\rangle &= \left(\frac{\epsilon_{\mu}s_{\nu}J_{1}^{\mu\nu}}{s\cdot k_{1}}-\frac{\epsilon_{\mu}s_{\nu}J_{n}^{\mu\nu}}{s\cdot k_{n}}\right)\left\langle \mathcal{V}_{1}...\mathcal{V}_{n}\right\rangle. 
\end{align}
Hence, we find that the leading and subleading terms in the soft limit of gluon amplitudes arise from the action of gauge transformations that are gauge analogues of supertranslations and superrotations. \\

We have thus seen the soft limits of tree-level graviton and gluon scattering amplitudes emerge as Ward identities for supertranslations and superrotations on $\scri$. The natural hamiltonian lift $h$ of diffeomorphisms of $\scri$ to the cotangent bundle $T^*\scri\cong\A$ allows us to define symmetry operators $Q_h$ inducing the action of the diffeomorphism on $\scri$ in the ambitwistor string. This in turn facilitates the identification of the leading and subleading terms in the soft limit of the integrated vertex operators as generators of supertranslations and superrotations on $\scri$, with the well-known soft terms emerging from the corresponding Ward identities.
 
\section{Four dimensional ambitwistor strings at $\scri$}
In the four dimensional case, adapting ambitwistor strings of \cite{Geyer:2014fka, Geyer:2014} to null infinity is perhaps more elegant, requiring no new coordinates.  This model uses the twistorial representation of ambitwistor space.  Twistor space is $\T=\C^{4|\cN}$ and we use coordinates 
$$
Z=(\mu^{\dot\alpha},\lambda_\alpha,\chi^a)\in \T\qquad \mbox{ and }\qquad W=(\tilde\lambda_{\dot\alpha},\tilde\mu^\alpha,\tilde \chi_a)\in\T^*
$$ 
where $\alpha=0,1$ and $\dot\alpha=\dot 0,\dot 1$ and $a=1\ldots \cN$.   Ambitwistor space is then represented as the quadric 
$$
\A=\{(Z,W)\in \T\times \T^*| Z\cdot W=0\}/\{Z\cdot\p_Z-W\cdot \p_W\} \,.
$$  
The symplectic potential in this representation is
$$
\Theta=\frac i2(Z\cdot \rd W-W\cdot \rd Z)\, .
$$

With homogeneous coordinates $(u,P_{\alpha\dot\alpha})$ on $\scri$ as before (using the spinorial decomposition of the vector index on $P$) we have that the projection from this representation of ambitwistor space to null infinity follows by setting \cite{Eastwood:1982} 
\be{}
u=-i\la \lambda \tilde\mu\ra \, , \qquad \tilde u = i[\tilde\lambda,\mu] \, , \qquad wp_{\alpha\dot\alpha}=P_{\alpha\dot\alpha}=\lambda_\alpha\tilde\lambda_{\dot\alpha}\, ,
\ee
where we have introduced the usual spinor helicity bracket notation to denote spinor contractions, $\la \lambda \tilde\mu\ra:=\lambda_\alpha \tilde\mu^\alpha\ra$ etc..  The spinorial representation here explicitly solves the constraint $P^2=0$, but, working without supersymmetry,  we see that $u=\tilde u$ is the constraint $Z\cdot W=0$. 

\subsection{Symmetries and hamiltonians}
Poincar\'e generators and supertranslations can easily be adapted to act on this ambitwistor space.   Indeed conformal motions $E^I_J\in \SU(2,2|\cN)$ are generated by $Z^JE_J^IW_I$.  The hamiltonian for the supertranslations $\delta u=f(\lambda,\tilde\lambda)$ with $f$ of weight $(1,1)$ in this model is simply $f$ itself as it induces the transformation
$$
\delta \tilde \mu^\alpha=i\frac {\p f}{\p\lambda_\alpha}\, , \quad \mbox{so} \quad \delta u =\lambda_\alpha \frac {\p f}{\p\lambda_\alpha}=f\,,
  $$
with the latter equality following by homogeneity.  Superrotations can similarly be taken to be those transformations generated by hamiltonians $H_r$ of weight $(1,1)$ that are linear in $(\mu,\chi)$ and in $(\tilde \mu,\tilde\chi)$ but have more complicated dependence in $(\lambda,\tilde\lambda)$, which will then of necessity include poles. These Poisson commute with $Z\cdot W$ on $Z\cdot W=0$ as they have weight $(1,1)$.  Thus we obtain
\be{}
H_r=[\mu, r]+\la\tilde\mu,\tilde r\ra\, , 
\ee
for $\tilde r_\alpha $  and $r_{\dot\alpha}$ respectively weight $(0,1)$ and $(1,0)$ functions of $(\lambda,\tilde\lambda)$.  These are linear functions respectively of $\tilde \lambda$ or $\lambda$ for ordinary rotations or dilations but for superrotations will be allowed to have poles and more general functional dependence on $(\lambda,\tilde\lambda)$. Below we will make the further requirement that 
$$
\frac{\p r_\alpha}{\p\lambda_\alpha}+ \frac{\p \tilde r_{\dot\alpha}}{\p \tilde\lambda_{\dot\alpha}} =\frac{\p^2 H_r}{\p Z^I\p W_I}=0
$$
which will ensure that we are working with $\SL(4)$ rather than $\GL(4)$.  

In order to incorporate Einstein gravity in the worldsheet model, we will introduce further coordinates $(\rho,\tilde\rho)\in \T\times \T^*$  of opposite statistics to $(Z,W)$ and perform the symplectic quotient by the following  further constraints
\begin{equation}\label{constraints}
Z\cdot\tilde \rho=W\cdot \rho=\rho\cdot \tilde\rho= \la Z\, \rho \ra = [W\, \tilde\rho]=0\, ,
\end{equation}
where $\la Z_1 Z_2\ra=\la \lambda_1 \lambda_2\ra$ and $[W_1\, W_2]=[\tilde\lambda _1\, \tilde \lambda_2]$.   In this model, the symplectic potential is 
$$
\Theta = \frac i2 \left (  Z\cdot \rd W -W\cdot \rd Z + \rho\cdot\rd \tilde\rho-\tilde \rho \rd \rho\right)\, .
$$ 

In order to extend the supertranslations and superrotations to this space, we need to extend the above Hamiltonians so that they commute with these constraints on the constraint submanifold.  It can be checked that this can be done automatically by taking the hamiltonians above and acting on them with $1+\rho \cdot \p_Z \tilde \rho\cdot \p_W$.  Thus for supertranslations we obtain the extensions
$$
\left(1+\rho \cdot \p_Z \tilde \rho\cdot \p_W\right)H_f=f + \rho^I \tilde\rho_J\frac{\p^2 f}{\p Z^I\p W_J}\, ,
$$
and for superrotations we get
$$
\left( 1+\rho \cdot \p_Z \tilde \rho\cdot \p_W\right)H_r=[\mu, r]+\la\tilde\mu,\tilde r\ra +  \rho^I \tilde\rho_J\frac{\p^2 ([\mu, r]+\la\tilde\mu,\tilde r\ra)}{\p Z^I\p W_J}\, .
$$

\subsection{The string model}
As before we base the action on the symplectic potential so that the
Poisson brackets will be reflected in the OPE.  For Yang-Mills we use
a model based on fields $(Z,W)$ on a Riemann surface $\Sigma$
that take values in $\T\times \T^*\otimes \kappa^{1/2}$ in which we
gauge the constraint $Z\cdot W=0$
$$
S=\int Z\cdot \dbar W-W\cdot \dbar Z +a Z\cdot W + e T
$$
where $a\in \Omega^{0,1}(\Sigma)$ is a Lagrange multiplier for the
constraint $Z\cdot W=0$, and as before $e\in\Omega^{0,1}\otimes
T\Sigma$ with  $T= Z\cdot \p W-W\cdot \p Z$, see \cite{Geyer:2014fka, Geyer:2014} for more detail.

For Yang-Mills, we introduce integrated vertex operators for both self-dual
and anti-self dual fields as 
\begin{eqnarray}{}
\cV_p&=&\int_\Sigma \frac{\rd s_p}{s_p} \bar\delta^2(\lambda_p-s_p\lambda
(\sigma_p)) \; \e^{is_p
  [\mu(\sigma_p) \tilde\lambda_p]}  j\cdot t_p \nonumber \\
\widetilde \cV_i&=&\int_\Sigma \frac{\rd s_i}{s_i} \bar\delta^2(\tilde
\lambda_i-s_i\tilde \lambda (\sigma_i)) \;  \e^{is_i
  \la\tilde \mu(\sigma_i) \lambda_i\ra}  j\cdot t_i  \label{VOym4d}
\end{eqnarray}
where $t_i$ is  a Lie algebra element and $j$ some current algebra on
$\Sigma$.\\

The gravity sector of the above model is the Berkovits-Witten
non-minimal version of conformal supergravity \cite{Berkovits:2004jj}.  
For Einstein gravity we must also incorporate the $(\rho,\tilde\rho)$ system described above to give the main matter action 
\begin{equation}
S=\int Z\cdot \dbar W-W\cdot \dbar Z + \tilde \rho\cdot\dbar\rho-\rho\cdot \dbar \tilde \rho\, .\\
\end{equation}
We gauge all the currents 
$$
K_a=\left(Z\cdot W, \rho\cdot \tilde \rho, Z\cdot \tilde \rho, W\cdot \rho, \la Z \rho\ra,  [W\, \tilde \rho]\right)
$$  
and gauge fixing setting all the gauge fields to zero, we are left with corresponding ghosts $(\beta_a,\gamma^a)$ and the usual BRST operator  
$$
Q_{BRST}=\oint cT+ \gamma^aK_a -\frac i2 \beta_a \gamma^b\gamma^c  C^a_{bc}\, ,
$$
where $C^a_{bc}$ are the structure constants of the current algebra $K_a$.

As in Yang-Mills, the pull-back from twistor space and dual twistor space leads to vertex operators for self-dual and anti self-dual fields \cite{Geyer:2014fka}, 
\begin{align}\label{VOgravity4d}
\cV_p&=\int_\Sigma \left(1+\rho \cdot \p_Z \tilde \rho\cdot \p_W\right)\frac{\rd s_p}{s_p^3} \bar\delta^2(\lambda_p-s\lambda
(\sigma_p)) \, [\tilde\lambda(\sigma_p) \, \tilde \lambda_p]   \, \e^{is_p
  [\mu(\sigma_p) \tilde\lambda_p]} \,,  \nonumber \\
\wt{\cV_i}&=\int_\Sigma \left(1+\rho \cdot \p_Z \tilde \rho\cdot \p_W\right)\frac{\rd s_i}{s_i^3} \bar\delta^2(\tilde\lambda_i-s\tilde\lambda
(\sigma_i)) \, \la\lambda(\sigma_i) \,  \lambda_i\ra   \, \e^{is_i
  \la\tilde\mu(\sigma_i) \lambda_i\ra}   \,.
\end{align}
It is easily seen that the integrated vertex operators defined as above agree with the original definition in \cite{Geyer:2014fka}.\\

The amplitude calculations for both Yang-Mills and gravity then reduce trivially to those of the original four-dimensional ambitwistor string \cite{Geyer:2014fka, Geyer:2014}, thus yielding the expected scattering amplitudes. As in higher dimensions, the correlation function will be evaluated by incorporating the exponentials of the vertex operators into the off-shell action. For $k$ insertions of $\wt\cV$ and $n-k$ insertions of $\cV$, the equations of motion determin $\lambda(\sigma)$ and $\tilde{\lambda}(\sigma)$ to be
\begin{equation}
 \lambda(\sigma)=\sum_{i=1}^k \frac{s_i\lambda_i}{\sigma-\sigma_i}, \qquad \tilde\lambda(\sigma)=\sum_{p=k+1}^n \frac{s_p\lambda_p}{\sigma-\sigma_p}\,.
\end{equation}

As in the higher dimensional case, the ambitwistor string theory naturally incorporates the geometry encoded in the Poisson structure via the singular part of the OPE, due to the construction based on the symplectic potential. The discussion of section \ref{symmetry} is therefore directly applicable in the four-dimensional case; for any hamiltonian $h$ generating a symplectic diffeomorphism on $\A$, the attribute that it preserves the symplectic potential and has thus the correct weights in the fields allows us to define a corresponding symmetry operator $Q_h$. In particular, the Hamiltonians for the BMS transformations discussed above will lead to operators inducing the action of the diffeomorphism of $\scri$ in the ambitwistor string.

\subsection{Soft limits}
\subsubsection*{Yang-Mills in 4d}
Following the same outline as in higher dimensions, we will again expand the integrated vertex operators \eqref{VOym4d} in the soft gluon limit to show that the leading and subleading terms correspond to generators of supertranslations and superrotations.\\

The $s$-integrals occuring in the Yang-Mills vertex operators can be performed explicitly against one of the delta
functions with a choice of reference spinors $\xi_\alpha$ or
$\tilde \xi_{\dot \alpha}$
 to give $s_s= \la \xi \, \lambda_s\ra/\la \xi \,
\lambda(\sigma_s)\ra$ in the first case and its tilde'd version in the
second.  For $\cV$ this leads to 
\begin{eqnarray}{}
\cV^{ym}_s&=&\int_\Sigma  \frac{\la \xi \, \lambda(\sigma_s)\ra}{\la \xi \,
  \lambda_s\ra }\bar\delta(\la \lambda_s\, \lambda (\sigma_s)\ra)\;
\exp \left(i
 \frac{\la \xi \, \lambda_s\ra [\mu(\sigma_s) \tilde\lambda_is}{\la \xi \,
 \lambda(\sigma_s)\ra }  \right)\;
J\cdot t_s \nonumber \\
&=&\oint
\frac{\la \xi \, \lambda(\sigma)\ra}{\la \xi \, \lambda_s\ra 
\la \lambda_s\, \lambda (\sigma_s)\ra }
\exp \left(i 
 \frac{\la \xi \, \lambda_s\ra [\mu(\sigma_s) \tilde\lambda_s]}{\la \xi \,
 \lambda(\sigma_s)\ra }  \right)\; 
J\cdot t_s \nonumber\\
&=& \cV^{ym,0}_s+\cV^{ym,1}_s +\cV^{ym,2}_s+\ldots
\end{eqnarray}
where, as before, in the second line we have used the fact that
\be{deltabar}
\bar\delta(\la \lambda_s\, \lambda (\sigma_s)\ra )=\dbar \frac1{2\pi i\la
\lambda_s\, \lambda (\sigma_s)\ra }
\ee
 and reduced the integral to a
contour integral around $\la \lambda_s\, \lambda (\sigma_s)\ra =0$.  In
the last line we are expanding the exponential in the soft gluon limit
$\lambda_s\tilde\lambda_s\rightarrow 0$.  We obtain
\begin{eqnarray}
\cV_s^{ym,0}&=&\oint \rd \sigma_s
\frac{\la \xi \, \lambda(\sigma_s)\ra}{\la \xi \, \lambda_s\ra 
\la \lambda_s\, \lambda (\sigma_s)\ra }J\cdot t_s\,, 
\nonumber \\
V_s^{ym,1}&=&\oint 
\frac{i[\mu(\sigma_s) \tilde\lambda_s]}{
\la \lambda_s\, \lambda (\sigma)\ra } \; J\cdot t_s\,,
\nonumber \\
V_s^{ym,2}&=&\oint 
\frac{-\la \xi\, \lambda_s\ra [\mu(\sigma_s) \tilde\lambda_s]^2}{ \la \xi
  \, \lambda(\sigma_s)\ra 
\la \lambda_s\, \lambda (\sigma_s)\ra } \; J\cdot t_s\,.
\end{eqnarray}
These can be thought of as singular gauge transformations, the gauge analogues of the supertranslations and BMS motions in the gravitational case below.

The Ward identities for a soft gluon, hence with a single insertion of the charges generating those singular gauge transformations, directly give the leading and subleading terms of the soft gluon limit,
\begin{align}
\left\langle \widetilde{\mathcal{V}}_{1}...\widetilde{\mathcal{V}}_{k}\mathcal{V}_{k+1}...\mathcal{V}_{n}\cV_s^{ym,0}\right\rangle& =\frac{\left\langle 1n\right\rangle }{\left\langle s1\right\rangle \left\langle sn\right\rangle }\left\langle \widetilde{\mathcal{V}}_{1}...\widetilde{\mathcal{V}}_{k}\mathcal{V}_{k+1}...\mathcal{V}_{n}\right\rangle \,,\\
\left\langle \widetilde{\mathcal{V}}_{1}...\widetilde{\mathcal{V}}_{k}\mathcal{V}_{k+1}...\mathcal{V}_{n}\cV_s^{ym,1}\right\rangle& =\left(\frac{1}{\left\langle s1\right\rangle }\tilde{\lambda}_{s}\cdot\frac{\partial}{\partial\tilde{\lambda}_{1}}+\frac{1}{\left\langle ns\right\rangle }\tilde{\lambda}_{s}\cdot\frac{\partial}{\partial\tilde{\lambda}_{n}}\right)\times \nonumber\\
&\qquad\qquad\times\left\langle \widetilde{\mathcal{V}}_{1}...\widetilde{\mathcal{V}}_{k}\mathcal{V}_{k+1}...\mathcal{V}_{n}\right\rangle \,.
\end{align}

\subsubsection*{Einstein gravity in 4d}
In analogy to the discussion in Yang-Mills, we can identify the leading and subleading terms in the soft expansion of the integrated gravity vertex operators as generators of supertranslations and superrotations on $\scri$; with the corresponding Ward identities yielding the soft graviton contributions found by Cachazo and Strominger, \cite{Cachazo:2014fwa}.

Following through the same steps as before, we get
\begin{eqnarray}{}
\cV_s&=&\int_\Sigma \left(1+\rho \cdot \p_Z \tilde \rho\cdot \p_W\right)\frac{\rd s}{s^3} \bar\delta^2(\lambda_s-s\lambda
(\sigma_s)) \, [\tilde\lambda(\sigma_s) \, \tilde \lambda_s]   \, \e^{is
  [\mu(\sigma_s) \tilde\lambda_s]}   \nonumber \\
&=& \int_\Sigma \left(1+\rho \cdot \p_Z \tilde \rho\cdot \p_W\right) \bar\delta(\la \lambda_s\, \lambda (\sigma_s)\ra) \frac{\la \xi \, \lambda(\sigma)\ra^2 [\tilde\lambda(\sigma) \, \tilde \lambda_s]}{\la \xi \,  \lambda_s\ra^2 }  \, 
\e^{ \left(i 
 \frac{\la \xi \, \lambda_s\ra [\mu(\sigma) \tilde\lambda_s]}{\la \xi \,
 \lambda(\sigma_s)\ra }  \right)}\nonumber \\
&=&\oint    \left(1+\rho \cdot \p_Z \tilde \rho\cdot \p_W\right)  \frac{\la \xi \, \lambda(\sigma_s)\ra^2 [\tilde\lambda(\sigma_s) \, \tilde \lambda_s]}{\la \xi \,  \lambda_s\ra^2 \la \lambda_s\, \lambda (\sigma_s)\ra} \,
\e^{ \left(i 
 \frac{\la \xi \, \lambda_s\ra [\mu(\sigma_s) \tilde\lambda_s]}{\la \xi \,
 \lambda(\sigma_s)\ra }  \right)}
 \nonumber\\
&=& \cV^0_s+\cV^1_s +\cV^2_s+\ldots
\end{eqnarray}
where, as above,  to get to the second line we have performed the $s$-integrals  against one of the delta
functions with a choice of reference spinor $\xi_\alpha$ to find $s= \la \xi\, \lambda_s\ra/ \la \xi \, \lambda(\sigma_s)\ra$.  To get to the third line we  have again used $\bar\delta(\la \lambda_s\, \lambda(\sigma_s)\ra)=\dbar (1/\la \lambda_s\,\lambda(\sigma_s)\ra)$.  In the last line we are simply expanding out the exponential as before to find
\begin{eqnarray}\label{4dgravV}
\cV^0_s&=&\oint    \left(1+\rho \cdot \p_Z \tilde \rho\cdot \p_W\right)  \frac{\la \xi \, \lambda(\sigma_s)\ra^2 [\tilde\lambda(\sigma_s) \, \tilde \lambda_s]}{\la \xi \,  \lambda_s\ra^2 \la \lambda_s\, \lambda (\sigma_s)\ra} \, 
\nonumber\\
\cV^1_s&=& \oint  \left(1+\rho \cdot \p_Z \tilde \rho\cdot \p_W\right)  \frac{i\la \xi \, \lambda(\sigma_s)\ra [\tilde\lambda(\sigma_s) \, \tilde \lambda_s]  [\mu(\sigma_s) \tilde\lambda_s]}{\la \xi \,  \lambda_s\ra \la \lambda_s\, \lambda (\sigma_s)\ra} \, \nonumber\\
\cV^2_s&=& \oint  \left(1+\rho \cdot \p_Z \tilde \rho\cdot \p_W\right)  \frac{ [\tilde\lambda(\sigma_s) \, \tilde \lambda_s]  [\mu(\sigma_s) \tilde\lambda_s]^2}{ \la \lambda_s\, \lambda (\sigma_s)\ra} \, 
\end{eqnarray}
Again, these explicit expressions allow for an identification of $\cV^0_s$ as a supertranslation generator, and of $\cV^1_s$ as a superrotation. $\cV^2_s$ corresponds, as in higher dimensions, to the `square' of a superrotation. As terms in the soft expansion of the vertex operators, all of these contributions generate diffeomorphisms of $\A$, but only $\cV^0_s$ and $\cV^1_s$ can be seen to arise from Hamiltonian lifts of diffeomorphisms of $\scri$.

In this 4-dimensional context, the superrotations are conformal on the sphere; being holomorphic in this negative helicity case and anti-holomorphic for a positive helicity soft graviton. 

With the vertex operators defined above for momentum eigenstates, an insertion of a soft graviton leads to the following Ward identities or the generators of supertranslations and superrotations respectively;
\begin{align}
 &\left\langle\wt{\mathcal{V}}_{1}...\wt{\mathcal{V}}_{k}\mathcal{V}_{k+1}...\mathcal{V}_{n}\;\cV^{0}_s\right\rangle=\sum_{a=1}^n\frac{[as]\la \xi\,a\ra^2}{\la a\,s\ra\la \xi\, s\ra^2}\left\langle\wt{\mathcal{V}}_{1}...\wt{\mathcal{V}}_{k}\mathcal{V}_{k+1}...\mathcal{V}_{n}\right\rangle\,,\\
 &\left\langle \wt{\mathcal{V}}_{1}...\wt{\mathcal{V}}_{k}\mathcal{V}_{k+1}...\mathcal{V}_{n}\;\cV^{1}_s\right\rangle=\sum_{a=1}^n\frac{[a\,s]\la\xi\,a\ra}{\la a\,s\ra\la\xi\,s\ra}\tilde\lambda_s\cdot\frac{\p}{\p \tilde\lambda_a}\left\langle \wt{\mathcal{V}}_{1}...\wt{\mathcal{V}}_{k}\mathcal{V}_{k+1}...\mathcal{V}_{n}\right\rangle\,,
\end{align}
refer to appendix \ref{grav4} for more details on the calculation. These Ward identities can immediately be seen to be equivalent to the leading and subleading terms in the soft graviton limit. 

In the sub-subleading case, $\cV^2_s$ generates a diffeomorphism of ambitwistor space corresponding to the `square' of a rotation, but does not descend to $\scri$ itself. The corresponding Ward identity, however, yields at tree-level still the predicted sub-subleading soft graviton contribution,
\begin{equation}
 \left\langle \wt{\mathcal{V}}_{1}...\wt{\mathcal{V}}_{k}\mathcal{V}_{k+1}...\mathcal{V}_{n}\;\cV^{2}_s\right\rangle
 =\frac1{2}\sum_{a=1}^n \frac{[a\,s]}{\la a\,s\ra}\tilde{\lambda}_s^{\dot\alpha}\tilde{\lambda}_s^{\dot\beta}\frac{\p^2}{\p\tilde\lambda_a^{\dot\alpha}\p\tilde\lambda_a^{\dot\beta}}\left\langle \wt{\mathcal{V}}_{1}...\wt{\mathcal{V}}_{k}\mathcal{V}_{k+1}...\mathcal{V}_{n}\right\rangle\,.
\end{equation}

\subsection{Brief comparison to the  model of Adamo et.\ al.\ }

This four dimensional twistorial ambitwistor model is closely connected to the 2d CFT recently proposed by Adamo, Casali, and Skinner \cite{Adamo:2014yya}. Indeed, the ambitwistor string is a very flexible framework and one can take different coordinate realizations of the space of null geodesics, adding further variables and corresponding constraints  to bring out different structures or features.  This can lead to quite different realizations with different properties (as witnessed by the distinction between the 4d RNS model versus the twistoral one which does a better job of bringing out underlying conformal invariance and its breaking).  The general strategy of identifying the worldsheet action from a symplectic potential guarantees that hamiltonians will give rise to operators that can be realized in the worldsheet theory.

 Their  model also lives on a supersymmetric extension of the cotangent bundle of the complexification of null infinity.  This model uses a different presentation of the supersymmetry, but the main coordinates can be identified by identifying their action as arising from the symplectic potential $\Theta$ on $T^*\scri$.  Thus we see for example that in their coordinates the symplectic potential is 
$$
\Theta= w \rd u + \chi\rd \xi+ \nu^A\rd \lambda_A+ \tilde \nu^{\dot A} \rd \tilde \lambda_{\dot A} + \bar \psi^A \rd \psi_A + \bar{\tilde \psi}^{\dot A}\rd \tilde\psi_{\dot A}\, , \quad A=(\alpha, a)\, ,
$$
and here $a=1,\ldots ,4$ is an R-symmetry index corresponding to a representation of $\cN=8$ supersymmetry.  We can clearly identify 
$$
Z=(\lambda_A,i\tilde \nu^{\dot A})\, , \quad  W= (-i\nu^{\dot A},\tilde \lambda_{\dot A})\, , \quad \rho=(\psi_A,\bar{\tilde \psi}^{\dot A})\, ,\quad  \tilde \rho=(\bar{ \psi}^{ A},\tilde\psi_{\dot A})\, ,
$$ 
because the bosonic parts of $\lambda$ and $\tilde\lambda$ are geometrically identical to that of the original twistorial ambitwistor model, 
although the representation of supersymmetry is somewhat different to the usual one on twistor space.  There are additional variables, $(w,u)$ playing an identical role to the $(w,u)$ in the $d$-dimensional ambitwistor string model. These are associated with an additional constraint that can be used to eliminate them and that is similarly the case here.  They also introduce a further pair of fields, $(\chi,\xi)$.     One can readily identify  $Z\cdot W=\la Z \rho \ra=[W\tilde \rho]=0$ constraints of the 4d ambitwistor string model amongst the gaugings in the ACS model.   

There are nevertheless important distinctions.  Firstly, the vertex operators are quite distinct from ours, and secondly their formulae work with the worldsheet fields taking values in line bundles of more general degree (ours are taken to be spinors on the worldsheet) leading to a larger integral over moduli in the evaluation of scattering amplitudes.  However, the latter issue is not so significant as it is already the case that the twistorial ambitwistor string model also admits different choices of degree, although these yield the same answer \cite{Geyer:2014}.

\section{Conclusion and discussion}

The work of Strominger and collaborators has shown that (extended) BMS symmetries have important implications for gravitational scattering amplitudes. In particular, if the diagonal subgroup of $BMS^+ \otimes BMS^-$ is taken to be a symmetry of the gravitational S-matrix,  the associated Ward identity for such supertranslations gives Weinberg's soft graviton theorem. Furthermore, Cachazo and Strominger found subleading terms in the soft graviton theorem and conjectured that they arise from the Ward identity associated with superrotations. The results were subsequently extended to Yang-Mills theory and to arbitrary spacetime dimension. Cachazo and Strominger also found a sub-subleading term in the soft limit for graviton ampitudes, although such a term does not seem to be associated with a symmetry of null infinity. 

These results are explained here using ambitwistor string theory in arbitrary dimensions.  Ambitwistor space can be identified with $T^*\scri^\pm$ and so admits a canonical lift of BMS symmetries.  If one expands the vertex operator of a soft graviton in powers of the soft momentum, the leading and subleading terms correspond to supertranslation and superrotation generators, thus confirming the the conjectures of Cachazo and Strominger.  Furthermore, we find that higher order terms in the expansion correspond to an infinite series of new soft terms which are associated with more general diffeomorphisms of ambitwistor space, although no longer lifted from diffeomorphisms of null infinity. These ideas work perhaps most elegantly in the case of the four dimensionsal twistorial version of the ambitwistor string model which does not need to be extended in any way to connect with null infinity.  Here, unlike the case of arbitrary dimension, the superrotations are local conformal motions of the sphere.  There are then two types of vertex operators, holomorphic or antiholomorphic depending on the helicity, generating two copies of the Virasoro algebra using vertex operators (we emphasize that this is a target space symmetry not worldsheet).

A remarkable feature is that gravitational vertex operators in ambitwistor string theory always arise as generators of rather more general  symplectic diffeomorphisms of $\A$. 
That such diffeomorphisms should encode the gravitational field
 follows from the original ambitwistor constructions of LeBrun
 \cite{LeBrun:1983} in which the gravitational field is encoded in the
 deformed complex structure of ambitwistor space.  In a \v Cech
 description of ambitwistor space the data is encoded in
 diffeomorphisms of the patching functions of the manifold.   In their
 infinitesimal form that leads to our vertex operators, these must be
 generated by hamiltonians to 
 preserve the holomorphic symplectic potential and 2-form as described in
 \cite{Baston:1987av}.
Here they are obtained from diffeomorphisms of $T^*\scri$ and in fact
they can be understood as arising from the scattering of null
geodesics through the space-time following the scattering theory
calculations of \cite{Penrose:1972ia}.  

What is therefore suggested by this picture is that we can give a description of the full nonlinear ambitwistor space as being obtained by glueing together a flat space one  constructed from the complexification of $\scri^-$ to another constructed from the complexification of $\scri^+$.  This then specifies enough of the complex structure on ambitwistor space to determine the full gravitational field and its scattering.
The scattering of null geodesics is already a complicated object and to identify those that correspond to solutions to Einstein's equations seems rather daunting in a fully nonlinear regime. 
However, within ambitwistor string theory, this is somehow achieved
perturbatively, but nevertheless to all orders, as the scattering of
null geodesics determined by each fourier mode in the vertex operator determines the scattering of the gravitational field by explicit ambitwistor-string calculation.  It would be intriguing to
find a nonperturbative formulation of this correspondence.  
In the ambitwistor string theory this might be expressed in the form
of the structure of a curved beta-gamma systems along the lines of
\cite{Witten:2005px,Nekrasov:2005wg} with gluing determined by
diffeomorphism from $\scri^-$ to $\scri^+$ arising from the scattering
of null geodesics but pieced together from manageable ingredients as it is in the perturbative calculations. 

The analagous story for Yang-Mills is that vertex operators at null infinity correspond
to certain gauge transformations at  $T^*\scri$.  In its soft expansion, we obtain  gauge transformations
analagous to supertranslations at leading order and superrotations for the subleading terms.
This gives a realization in string theoretic terms of the Ambitwistor constructions of \cite{Witten:1978xx, Isenberg:1978kk,Witten:1985nt} in which Yang-Mlls fields are encoded in the complex structure of a holomorphic vector bundle over ambitwistor twistor space with the gauge transformations playing the role of patching functions.  

 Loop corrections in ambitwistor string theory have been studied in
\cite{Adamo:2013tsa} and it seems plausible that ambitwistor strings
will give the correct all-loop integrand for type II supergravity in
10 dimensions, although this remains speculative.  If so these ideas will apply directly in that context (and hence to its reductions) also.

\begin{center}
\textbf{Acknowledgements}
\end{center}

AL is supported by a Simons Postdoctoral Fellowship, YG by the EPSRC and the Mathematical Prizes fund and LM is supported by EPSRC grant number EP/J019518/1. 
We thank Dave Skinner for stimulating conversations.

\appendix 
\section{Details of the correlators with soft limits}

Integrated gluon and graviton vertex operators implement symplectic diffeomorphisms of $T^*\scri$ in the worldsheet ambitwistor string theory. We have seen explicitly how these vertex operators can be expanded in powers of the soft momentum, and identified the leading and subleading term as generators of supertranslations and superrotations on $\scri$ (and the corresponding gauge transformation analogues for Yang-Mills). In this appendix, we calculate the associated Ward identities, both in the $d$ dimensional model and the four dimensional twistorial model, and prove that they correspond at leading order to Weinberg's soft gluon and graviton theorems, and at subleading order to the gluon and graviton theorems derived in \cite{Cachazo:2014fwa,Casali:2014xpa}.

\subsection{Yang-Mills soft limits in $d$-dimensional model} \label{ymd}

\subsubsection*{Leading terms}

Let $\epsilon$ and $s$ be the polarization and momentum
of a soft gluon. If we expand the vertex operator in $s$, the leading
term corresponds to the generator of a singular gauge transformation that only
depends on $p$.  This is the gauge analogue of a  supertranslation and
we denote it by $\mathcal{V}_s^{ym,0}$:
\[
\mathcal{V}_s^{ym,0}=\frac{1}{2 \pi i}\oint d\sigma_{s}\frac{\epsilon\cdot
  p(\sigma_{s})}{s\cdot p\left(\sigma_{s}\right)}j(\sigma_{s})
\]
where $j(\sigma_{s})$ is the worldsheet current algebra contracted
with an
element of the corresponding Lie algebra. Since we are dealing with
color-stripped amplitudes, we will leave out generators of the Lie
algebra and simply take the single trace term when we take the
correlation function. 

Consider the correlator of a soft gluon with $n$ other gluons. This is given by
\begin{equation*}
\left\langle \mathcal{V}_{1}...\mathcal{V}_{n} \mathcal{V}_s^{ym,0} \right\rangle =\frac{1}{2 \pi i}\sum_{j=1}^{n}\left\langle \mathcal{V}_{1}...\mathcal{V}_{n}\oint_{\left|\sigma_{s}-\sigma_{j}\right|<\epsilon}d\sigma_{s}\frac{\left(\sigma_{n}-\sigma_{1}\right)}{\left(\sigma_{s}-\sigma_{1}\right)\left(\sigma_{n}-\sigma_{s}\right)}\frac{\epsilon\cdot p(\sigma_{s})}{s\cdot p\left(\sigma_{s}\right)}\right\rangle \label{eq:ym}
\end{equation*}
where $\epsilon \rightarrow 0$, we have used \eqref{p} and used the
current the single trace term in the current correlator to obtain a
Parke-Taylor denominator from which we have extracted the soft term.
As the soft gluon vertex operator approaches
one of the other vertex operators, we have 
\begin{equation}
\lim_{\sigma_{s}\rightarrow\sigma_{j}}\frac{\epsilon\cdot p(\sigma_{s})}{s\cdot p\left(\sigma_{s}\right)}=\frac{\epsilon\cdot k_{j}}{s\cdot k_{j}}.
\end{equation}
Plugging this into equation \eqref{eq:ym} and performing the contour
integral finally gives the leading order contribution to the soft limit,
\begin{equation}
\left\langle \mathcal{V}_{1}...\mathcal{V}_{n} \mathcal{V}_s^{ym,0}\right\rangle =\left(\frac{\epsilon\cdot k_{1}}{s\cdot k_{1}}-\frac{\epsilon\cdot k_{n}}{s\cdot k_{n}}\right)\left\langle \mathcal{V}_{1}...\mathcal{V}_{n}\right\rangle .
\end{equation}

\subsubsection*{Subleading terms}

Expanding the vertex operator further in $s$, the gauge analogue of
the superrotation generator corresponds the terms linear in $s$ 
\[
\mathcal{V}_s^{ym,1}=Q_R^{orbit}+Q_R^{spin}
\]
where 
\begin{align}
Q_{R}^{orbit}&=\frac{1}{2\pi i}\oint d\sigma_{s}\frac{iq(\sigma_{s})\cdot s\,\epsilon\cdot p(\sigma_{s})}{s\cdot p(\sigma_{s})}j(\sigma_{s})\\
Q_{R}^{spin}&=\frac{1}{2\pi i}\oint
d\sigma_{s}\frac{\epsilon\cdot\Psi(\sigma_{s})s\cdot\Psi(\sigma_{s})}{s\cdot
  p(\sigma_{s})}j(\sigma_{s})\, .
\end{align}
 Let's compute the correlator
of $Q_{R}$ with $n$ other vertex operators. If we focus only on
the delta functions in the other vertex operators, we can neglect
$Q_{R}^{spin}$, since the delta functions do not depend on fermionic
fields. Hence, we only need the following OPE: 
\begin{equation}
is\cdot v(\sigma_{s})\bar{\delta}\left(k_{j}\cdot P(\sigma_{j})\right)=\frac{k_{j}\cdot s}{\sigma_{s}-\sigma_{j}}\bar{\delta}^{(1)}\left(k_{j}\cdot P(\sigma_{j})\right)+...\label{eq:ope2}
\end{equation}
where $\bar{\delta}^{(n)}(x)=\left(\frac{\partial}{\partial x}\right)^{n}\left(x^{-1}\right)$, which follows from \eqref{eq:ope}. Focusing on the delta functions
of the vertex operators and using the above OPE, one easily finds
that 
\[
\left\langle \mathcal{V}_{1}...\mathcal{V}_{n}Q_{R}^{orbit}\right\rangle =\frac{1}{2\pi i}\int d^{2n}\sigma\oint d\sigma_{s}\frac{\epsilon\cdot P\left(\sigma_{s}\right)}{s\cdot P(\sigma_{s})}\frac{\sigma_{n1}}{\sigma_{s1}\sigma_{ns}}\sum_{j=1}^{n}\frac{s\cdot k_{j}}{\sigma_{sj}}\bar{\delta}_{j}^{(1)}\Pi_{a=1,a\neq j}^{n}\bar{\delta}_{a}I_{n}
\]
where $\sigma_{ij}=\sigma_{i}-\sigma_{j}$, $\bar{\delta}_{j}=\bar{\delta}\left(k_{j}\cdot P(\sigma_{j})\right)$,
and $I_{n}$ indicates that rest of the integrand does not depend
on $\sigma_{s}$. Note that this integral is precisely equation (19)
of \cite{Schwab:2014xua}. Following the calculations of this paper, we can easily see that this will indeed correspond to the subleading soft limit terms $S^{(1)}$, with the derivatives taken to act exclusively on the scattering equations obtained from the momentum eigenstates in the vertex operators.\\

To obtain the full subleading soft factors, we will have to include all contributions from the correlation function $\left\langle \mathcal{V}_{1}...\mathcal{V}_{n}Q_{R}^{orbit}\right\rangle$, as well as additional contributions from $Q_R^{spin}$. In particular, we find that
\[
\begin{split}
 &\left\langle \mathcal{V}_{1}...\mathcal{V}_{n}Q_{R}^{orbit}\right\rangle \\
 &\qquad=\frac{1}{2\pi i}\int \frac{d^{n}\sigma}{\vol \SL(2)}\oint d\sigma_{s}\frac{\epsilon\cdot P\left(\sigma_{s}\right)}{s\cdot P(\sigma_{s})}\frac{\sigma_{n1}}{\sigma_{s1}\sigma_{ns}}\\
 &\qquad\qquad\left(\sum_{j=1}^{n}\frac{s\cdot k_{j}}{\sigma_{sj}}\bar{\delta}_{j}^{(1)}\Pi_{a=1,a\neq j}^{n}\bar{\delta}_{a}\frac{\text{Pf}(M^{(n)})}{\prod_{b}\sigma_{b,b+1}}+\sum_{a=1}^n \frac{s\cdot \epsilon_a}{\sigma_{sa}}\Pi_b'\bar{\delta}_b\frac{\text{Pf}({M^{(n)}}^{a,a+n}_{a,a+n})}{\prod_{b}\sigma_{b,b+1}}\right)\,,
\end{split}
\]
where we denote the CHY matrix obtained from $n$ vertex operator insertions by $M^{(n)}$. Note especially that this does not contain any data of the soft gluon. As mentioned above, additional contributions to the orbital part of the subleading soft limit (in addition to the spin contribution) will originate from the correlation function involving $Q_R^{spin}$;
\[
 \begin{split}
  &\left\langle \mathcal{V}_{1}...\mathcal{V}_{n}Q_{R}^{spin}\right\rangle \\
 &\qquad=\frac{1}{2\pi i}\int \frac{d^{n}\sigma}{\vol \SL(2)}\frac{1}{\prod_{b}\sigma_{b,b+1}}\oint d\sigma_{s}\frac1{s\cdot P(\sigma_{s})}\frac{\sigma_{n1}}{\sigma_{s1}\sigma_{ns}}\\
 &\qquad\qquad\sum_{a,b}(-1)^{a+b}\Bigg(\frac{\epsilon\cdot k_a}{\sigma_{sa}}\frac{s\cdot\epsilon_b}{\sigma_{sb}}\text{Pf}({M^{(n)}}^{a,b+n}_{a,b+n})
 -\frac{s\cdot k_a}{\sigma_{sa}}\frac{\epsilon\cdot \epsilon_b}{\sigma_{sb}}\text{Pf}({M^{(n)}}^{a,b+n}_{a,b+n})\\
 &\qquad\qquad+\frac{\epsilon\cdot k_a}{\sigma_{sa}}\frac{s\cdot k_b}{\sigma_{sb}}\text{Pf}({M^{(n)}}^{a,b}_{a,b})-\frac{\epsilon\cdot \epsilon_a}{\sigma_{sa}}\frac{s\cdot\epsilon_b}{\sigma_{sb}}\text{Pf}({M^{(n)}}^{a+n,b+n}_{a+n,b+n})\Bigg)\,.
 \end{split}
\]
A closer look at the structure and origin of these terms already indicates how to match them to the contributions to the subleading soft limits found in \cite{Schwab:2014xua}. Recall from the original ambitwistor string \cite{Mason:2013sva} that in the correlation functions, the fermionic fields $\Psi$ give rise to the Pfaffians, with the diagonal terms $C_{aa}$ coming from the contributions $\epsilon\cdot P(\sigma)$. An insertion of $Q_{R}^{orbit}$ will therefore contribute the subleading soft limits, where the derivative is taken to act on the scattering equations, as well as an additional term due to the appearance of the soft gluon in the diagonal terms of the matrix $C$. The charge $Q_{R}^{spin}$, on the other hand, will give the remaining contributions of the soft particle in the Pfaffian, as well as the spin contribution $J_{spin,a}^{\mu\nu}=\epsilon_{a}^{[\mu} k_{a}^{\nu]}$, stemming from the double contractions where both soft gluon $\Psi_s$ fields are contracted to the fields $\Psi_a$ of {\it one} external gluon $a$. Combining these terms and following the manipulations described in \cite{Schwab:2014xua}, one then finds the subleading soft limit,
\begin{equation}
  \left\langle \mathcal{V}_{1}\dots\mathcal{V}_{n}\mathcal{V}_s^{ym,1}\right\rangle= \left(\frac{\epsilon_{\mu}s_{\nu}J_{1}^{\mu\nu}}{s\cdot k_{1}}-\frac{\epsilon_{\mu}s_{\nu}J_{n}^{\mu\nu}}{s\cdot k_{n}}\right)\left\langle \mathcal{V}_{1}...\mathcal{V}_{n}\right\rangle 
\end{equation}
where $J_a^{\mu\nu}=J_{orb,a}^{\mu\nu}+J_{spin,a}^{\mu\nu}$, with $J_{orb,a}^{\mu\nu}=k_{a}^{[\mu}\frac{\partial}{\partial k_{a,\nu]}}$ and $J_{spin,a}^{\mu\nu}=\epsilon_{a}^{[\mu} k_{a}^{\nu]}$.

\subsection{Gravity soft limits in $d$-dimensional model} \label{gravd}

\subsubsection*{Leading terms}
For a soft graviton $s$, we are interested in computing the Ward identitiy associated to the leading order term $\cV_{s}^0$ in the soft expansion of the vertex operator. As we have seen above, this corresponds to a supertranslation on $\scri$. With one insertion of $\cV_{s}^0$, the correlator becomes
\begin{equation*}
\left\langle \mathcal{V}_{1}...\mathcal{V}_{n} \cV_{s}^0\right\rangle =\frac{1}{2\pi i}\sum_{j=1}^{n}\left\langle \mathcal{V}_{1}...\mathcal{V}_{n}\oint_{\left|\sigma_{s}-\sigma_{j}\right|<\epsilon}d\sigma_{s}\frac{(\epsilon\cdot p(\sigma_{s}))^2}{s\cdot p\left(\sigma_{s}\right)}\right\rangle \label{eq:gr}
\end{equation*}
where $\epsilon \rightarrow 0$ and $p(\sigma)$ is given in equation \eqref{p}. When the soft graviton vertex operator approaches one of
the other vertex operators, from \eqref{p} we have 
\[
\lim_{\sigma_{s}\rightarrow\sigma_{j}}\frac{(\epsilon\cdot p(\sigma_{s}))^2}{s\cdot p\left(\sigma_{s}\right)}=\frac{(\epsilon\cdot k_{j})^2}{s\cdot k_{j}\left(\sigma_{s}-\sigma_{j}\right)}.
\]
Plugging this into equation \eqref{eq:gr} and performing the contour
integral yields the Weinberg soft graviton theorem,
\begin{equation}
\left\langle \mathcal{V}_{1}...\mathcal{V}_{n} \cV_{s}^0\right\rangle =\left(\sum_{j=1}^{n}\frac{(\epsilon\cdot k_j)^2}{s\cdot k_{j}}\right)\left\langle \mathcal{V}_{1}...\mathcal{V}_{n}\right\rangle. 
\end{equation}

\subsubsection*{Subleading terms}
Expanding the soft graviton vertex operator further in $s$, we obtain a term $\cV^1_s$ linear in $s$ which corresponds to the generator of a supertranslation on $\scri$. Note that $\cV^1_s$ is made out of $r_{\mu\nu}J^{\mu\nu}$ which breaks up into an orbital part $q^{[\mu}p^{\nu]}$ and spin part $\Psi_r^\mu\Psi_r^\nu$:
\[
\mathcal{V}_s^{1}=Q_R^{orbit}+Q_R^{spin},
\]
where the orbital and spin contributions are given by
\begin{align}
Q_{R}^{orbit}&=\frac{1}{2\pi i}\oint d\sigma_{s}\frac{i \epsilon\cdot p l^{[\mu}\epsilon^{\nu]} q(\sigma_{s})_\mu p(\sigma_{s})_\nu}{s\cdot p(\sigma_{s})}\\
Q_{R}^{spin}&=\frac{1}{2\pi i}\oint d\sigma_{s}\frac{\epsilon\cdot p(\sigma_{s})s\cdot\Psi_{2}(\sigma_{s})\epsilon\cdot\Psi_{2}(\sigma_{s})+(1\leftrightarrow2)}{s\cdot p(\sigma_{s})}.
\end{align}
The correlation functions involving these vertex operators are computed using the OPE \eqref{eq:ope}. 
Related calculations have been performed in detail in \cite{Schwab:2014xua} to compute subleading soft
limits. There, the authors focus on the soft limits of the
delta functions in the CHY formulae, which contributes to
the orbital part of the subleading soft limit. The remainder of the orbital part and the spin part of the
subleading soft limit then comes from analyzing the soft limits of
the Pfaffians. Similarly, when we compute the correlation functions
of $\mathcal{V}_s^{1}$ with other vertex operators, we will first focus on the contractions involving the delta
functions of the other vertex operators. This will allow us to make contact with the calculations in \cite{Schwab:2014xua} to demonstrate that $Q_{R}^{orbit}$ indeed generates the correct contributions to the orbital part
of the subleading soft limit. One can then show that $Q_{R}^{spin}$ generates the spin part of the subleading soft limit, as well as the missing contributions to the orbital part. $\mathcal{V}_s^{1}=Q_R^{orbit}+Q_R^{spin}$ will therefore generate the full subleading soft gluon or graviton contribution as discussed in \cite{Cachazo:2014fwa}.

In compute the correlator of $Q_{R}$ with $n$ other vertex operators, we will focus first only on the delta functions in the other vertex operators,
and neglect $Q_{R}^{spin}$. Furthermore, using equation \ref{eq:ope2},
one finds that
\[
\left\langle \mathcal{V}_{1}...\mathcal{V}_{n}Q_{R}^{orbit}\right\rangle =\frac{1}{2\pi i}\int d^{2n}\sigma\oint d\sigma_{s}\frac{\epsilon_{1}\cdot P(\sigma_{s})\epsilon_{2}\cdot P(\sigma_{s})}{s\cdot P(\sigma_{s})}\sum_{j=1}^{n}\frac{s\cdot k_{j}}{\sigma_{sj}}\bar{\delta}_{j}^{(1)}\Pi_{a=1,a\neq j}^{n}\bar{\delta}_{a}I_{n}
\]
where we use notation defined in the previous subsection. Note that this integral is precisely
equation 23 of \cite{Schwab:2014xua}. Again, the remaining correlation function,
\[
\left\langle \mathcal{V}_{1}...\mathcal{V}_{n}Q_{R}^{spin}\right\rangle\,,
\]
can be calculated along simlar lines as in Yang-Mills, described in appendix \ref{ymd}. Following the manipulations outlined in \cite{Schwab:2014xua}, we find indeed the subleading soft graviton limit derived in \cite{Cachazo:2014fwa}
\begin{equation}
\left\langle \mathcal{V}_{1}...\mathcal{V}_{n}\mathcal{V}_s^{1}\right\rangle =\sum_{a=1}^{n}\frac{\epsilon_{\mu\nu}k_{a}^{\mu}s_{\lambda}J_{a}^{\lambda\nu}}{s\cdot k_{a}}\left\langle \mathcal{V}_{1}...\mathcal{V}_{n}\right\rangle 
\end{equation}
where $\epsilon^{\mu\nu}=\epsilon_{1}^{(\mu}\epsilon_{2}^{\nu)}$
and $J_{a}^{\mu\nu}$ was defined in appendix \ref{ymd}.

\subsection{Yang-Mills soft limits in the twistorial model} \label{ym4}

\subsubsection*{Leading terms}

The action of the worldsheet model for the ambitwistor string is based on the symplectic potential of $\A$, and the singular parts of the OPE of operators in the ambitwistor string is thus given by the Poisson structure on $\A=T^*\scri$. In calculating the soft limits in the twistorial model, the following OPE's of fields in the ambitwistor string will be useful:
\begin{equation}
\lambda_{\alpha}(z)\tilde{\mu}^{\beta}(w)=\frac{\delta_{\alpha}^{\beta}}{z-w}+...,\,\,\,\tilde{\lambda}_{\dot{\alpha}}(z)\mu^{\dot{\beta}}(w)=\frac{\delta_{\dot{\alpha}}^{\dot{\beta}}}{z-w}+...\label{eq:sope}
\end{equation}

Expanding an integrated gluon vertex operator in the soft momentum,
the leading term is given by

\[
\mathcal{V}^{ym,0}_s=\frac{1}{2\pi i}\oint d\sigma_{s}\frac{\left\langle \xi\lambda\left(\sigma_{s}\right)\right\rangle }{\left\langle \xi\,s\right\rangle \left\langle s\,\lambda\left(\sigma_{s}\right)\right\rangle }j\left(\sigma_{s}\right)
\]
where $l=\lambda_{s}\tilde{\lambda}_{s}$ is the soft momentum,
$\xi_{\alpha}$ is a reference spinor, and 
\begin{equation}
\lambda(\sigma)=\sum_{i=1}^{k}\frac{s_{i}\lambda_{i}}{\sigma-\sigma_{i}}.\label{eq:lambdas}
\end{equation}
Let us compute the correlator of $\mathcal{V}^{ym,0}_s$ with $k$ negative helicity
vertex operator $\wt{\mathcal{V}}$ and $n-k$ positive helicity
vertex operators $\mathcal{V}$:
\begin{equation}
\left\langle \wt{\mathcal{V}}_{1}...\wt{\mathcal{V}}_{k}\mathcal{V}_{k+1}...\mathcal{V}_{n}\mathcal{V}^{ym,0}_s\right\rangle =\frac{1}{2\pi i}\frac{1}{\left\langle \xi\,s\right\rangle }\left\langle \wt{\mathcal{V}}_{1}...\wt{\mathcal{V}}_{k}\mathcal{V}_{k+1}...\mathcal{V}_{n}\oint d\sigma_{s}\frac{\sigma_{n1}}{\sigma_{ns}\sigma_{s1}}\frac{\left\langle \xi\,\lambda\left(\sigma_{s}\right)\right\rangle }{\left\langle s\,\lambda\left(\sigma_{s}\right)\right\rangle }\right\rangle .\label{eq:ymqt}
\end{equation}
Note from equation \ref{eq:lambdas} that
\[
\lim_{\sigma_{s}\rightarrow\sigma_{1}}\frac{\left\langle \xi\,\lambda\left(\sigma_{s}\right)\right\rangle }{\left\langle s\,\lambda\left(\sigma_{s}\right)\right\rangle }=\frac{\left\langle \xi\,1\right\rangle }{\left\langle s\,1\right\rangle }.
\]
Furthermore, on the support of the delta functions in $\mathcal{V}_{n}$,
we have 
\[
\lim_{\sigma_{s}\rightarrow\sigma_{n}}\frac{\left\langle \xi\,\lambda\left(\sigma_{s}\right)\right\rangle }{\left\langle s\,\lambda\left(\sigma_{s}\right)\right\rangle }=\frac{\left\langle \xi\,n\right\rangle }{\left\langle s\,n\right\rangle }.
\]
Hence, when we evaluate the contour integral in \eqref{eq:ymqt},
the residues at $\sigma_{s}=\sigma_{1}$ and $\sigma_{s}=\sigma_{n}$
give the soft graviton contribution to leading order
\[
\left\langle \wt{\mathcal{V}}_{1}...\wt{\mathcal{V}}_{k}\mathcal{V}_{k+1}...\mathcal{V}_{n}\mathcal{V}^{ym,0}_s\right\rangle =\frac{\left\langle 1\,n\right\rangle }{\left\langle s\,1\right\rangle \left\langle s\,n\right\rangle }\left\langle \wt{\mathcal{V}}_{1}...\wt{\mathcal{V}}_{k}\mathcal{V}_{k+1}...\mathcal{V}_{n}\right\rangle 
\]
where we have used the Schouten identity.

\subsubsection*{Subleading terms}

Expanding the gluon vertex operator further to first order in the soft momentum
gives
\[
\mathcal{V}^{ym,1}_s=\frac{1}{2\pi i}\oint d\sigma_{s}\frac{\left[\mu\left(\sigma_{s}\right)s\right]}{\left\langle s\,\lambda(\sigma_{s})\right\rangle }J\left(\sigma_{s}\right).
\]
Note that there is subtlety in defining this operator, since the equations
of motion for the $\tilde{\lambda}$ field imply that $\mu=0$. On
the other hand, $\mu$ will have nonzero contractions with the $\tilde{\lambda}$
fields which appear in the delta functions of other vertex operators,
so correlation functions of $\mathcal{V}^{ym,1}_s$ will be non-vanishing. In particular,
from \eqref{eq:sope}, we see that 
\[
\left[\mu\left(\sigma_{s}\right)s\right]\bar{\delta}^{2}\left(\tilde{\lambda}_{i}-t_{i}\tilde{\lambda}\left(\sigma_{i}\right)\right)=\frac{1}{\sigma_{s}-\sigma_{i}}\tilde{\lambda}_{s}\cdot\frac{\partial}{\partial\tilde{\lambda}\left(\sigma_{i}\right)}\bar{\delta}^{2}\left(\tilde{\lambda}_{i}-t_{i}\tilde{\lambda}\left(\sigma_{i}\right)\right)+...
\]
where the ellipses denote non-singular terms. 
The subleading contribution to the soft gluon will arise from the correlator of $\mathcal{V}^{ym,1}_s$ with $k$ negative helicity
vertex operator $\wt{\mathcal{V}}$ and $n-k$ positive helicity
vertex operators $\mathcal{V}$:
\begin{equation*}
\left\langle \tilde{\mathcal{V}}_{1}...\tilde{\mathcal{V}}_{k}\mathcal{V}_{k+1}...\mathcal{V}_{n}\mathcal{V}^{ym,1}_s\right\rangle =\frac{1}{2\pi i}\int d^{2n}\sigma\oint d\sigma_{s}\frac{1}{\left\langle s\lambda(\sigma_{s})\right\rangle }\frac{\sigma_{n1}}{\sigma_{ns}\sigma_{s1}}\sum_{i=1}^{k}\frac{1}{\sigma_{si}}\tilde{\lambda}_{s}\cdot\frac{\partial}{\partial\tilde{\lambda}\left(\sigma_{i}\right)}I_{n}\label{eq:ymr}
\end{equation*}
where $I_{n}$ indicates that the rest of the integrand does not depend
on $\sigma_{s}$. Noting that
\[
\lim_{\sigma_{s}\rightarrow\sigma_{1}}\frac{1}{\left\langle s\lambda(\sigma_{s})\right\rangle }=\frac{\sigma_{s1}}{s_{1}\left\langle s1\right\rangle },
\]
the residue at $\sigma_{s}=\sigma_{1}$ gives to
\[
\frac{1}{\left\langle s1\right\rangle }\tilde{\lambda}_{s}\cdot\frac{\partial}{\partial\tilde{\lambda}_{1}}\left\langle \wt{\mathcal{V}}_{1}...\wt{\mathcal{V}}_{k}\mathcal{V}_{k+1}...\mathcal{V}_{n}\right\rangle .
\]
Furthermore, the residue at $\sigma_{s}=\sigma_{n}$ corresponds to
\[
\int d^{2n}\sigma\frac{1}{\left\langle \lambda(\sigma_{n})s\right\rangle }\sum_{i=1}^{k}\frac{1}{\sigma_{ni}}\tilde{\lambda}_{s}\cdot\frac{\partial}{\partial\tilde{\lambda}\left(\sigma_{i}\right)}I_{n}=\frac{1}{\left\langle ns\right\rangle }\tilde{\lambda}_{s}\cdot\frac{\partial}{\partial\tilde{\lambda}_{n}}\left\langle \wt{\mathcal{V}}_{1}...\wt{\mathcal{V}}_{k}\mathcal{V}_{k+1}...\mathcal{V}_{n}\right\rangle 
\]
where we noted that on the suppport of the delta functions in $\mathcal{V}_{n}$, we have
$\left\langle \lambda(\sigma_{n})s\right\rangle =\left\langle ns\right\rangle /s_{n}$.
Hence, we find that the correlator in equation \ref{eq:ymr} reduces
to the subleading soft gluon contribution from \cite{Casali:2014xpa}
\[
\left\langle \tilde{\mathcal{V}}_{1}...\tilde{\mathcal{V}}_{k}\mathcal{V}_{k+1}...\mathcal{V}_{n}\mathcal{V}^{ym,1}_s\right\rangle =\left(\frac{1}{\left\langle s1\right\rangle }\tilde{\lambda}_{s}\cdot\frac{\partial}{\partial\tilde{\lambda}_{1}}+\frac{1}{\left\langle ns\right\rangle }\tilde{\lambda}_{s}\cdot\frac{\partial}{\partial\tilde{\lambda}_{n}}\right)\left\langle \tilde{\mathcal{V}}_{1}...\tilde{\mathcal{V}}_{k}\mathcal{V}_{k+1}...\mathcal{V}_{n}\right\rangle .
\]

\subsection{Gravity soft Limits in twistorial model} \label{grav4}

As we have seen above, the terms in the soft limit expansion of the integrated vertex operators for gravity correspond to generators for the symmetries of $\scri$, in particular we find generators of translations $\cV_s^0$ at leading order, and generators of superrotations $\cV_s^1$ at subleading order. By imposing the constraints \eqref{constraints}, the equations for the generators \eqref{4dgravV} can be simplified drastically. Moreover, in contraction with the vertex operators introduced above, the pieces $\rho_\alpha\frac{\p}{\p\lambda_\alpha}+\rho_A\frac{\p}{\p\chi_A}$ can be ignored, as there remains always at least one $\tilde\rho$ in one of the vertex operators, which causes the path integral to vanish. Keeping this in mind, the symmetry generators due to a soft graviton are given by
\begin{align}
\cV_s^0&=\frac1{2\pi i}\oint    \frac{\la \xi \, \lambda(\sigma_s)\ra^2 [\tilde\lambda(\sigma_s) \, \tilde \lambda_s]}{\la \xi \,  \lambda_s\ra^2 \la \lambda_s\, \lambda (\sigma_s)\ra} \,, \\
 \cV_s^1&= \frac{1}{2\pi}\oint  \left( \frac{\la \xi \, \lambda(\sigma_s)\ra [\tilde\lambda(\sigma_s) \, \tilde \lambda_s]  [\mu(\sigma_s) \tilde\lambda_s]}{\la \xi \,  \lambda_s\ra \la \lambda_s\, \lambda (\sigma_s)\ra}+\frac{\la \xi\,\lambda(\sigma_s)\ra[\rho\,\tilde\lambda_s][\tilde\rho\,\tilde\lambda_s]}{\la\xi\,\lambda_s\ra\la\lambda_s\,\lambda(\sigma_s)\ra}\right) \,,\\
 \cV_s^2&= \frac1{2\pi i}\oint  \left(  \frac1{2}\frac{ [\tilde\lambda(\sigma_s) \, \tilde \lambda_s]  [\mu(\sigma_s) \tilde\lambda_s]^2}{ \la \lambda_s\, \lambda (\sigma_s)\ra}+\frac{[\rho\,\tilde\lambda_s][\tilde\rho\,\tilde\lambda_s][\mu(\sigma_s)\,\tilde\lambda_s]}{\la\lambda_s\,\lambda(\sigma_s)\ra}\right) \, .
\end{align}

\subsubsection*{Leading terms}

In particular, we can investigate the Ward identity of the first order contribution of an integrated vertex operator in the soft limit, which we have identified with a charge associated to superrotations. For a soft graviton $s$, the superrotation generator is then given by
\begin{equation}
\cV_s^0=\frac1{2\pi i}\oint \rd \sigma_s\frac{\la \xi \, \lambda(\sigma_s)\ra^2 [s\,\tilde\lambda(\sigma_s)]}{\la \xi \,  s\ra^2 \la s\, \lambda (\sigma_s)\ra}.
\end{equation}
We are interested in the Ward identity 
\begin{equation*}
 \left\langle\cV_1\dots \cV_n\cV_s^0\right\rangle\,,
\end{equation*}
for momentum eigenstates, where the equations of motion determine $\lambda(\sigma)$ and $\tilde\lambda(\sigma)$ to be
\begin{equation*}
 \lambda(\sigma)=\sum_{i=1}^k \frac{s_i\lambda_i}{\sigma-\sigma_i}\,,\qquad \tilde\lambda(\sigma)=\sum_{p=k+1}^n \frac{s_p\tilde\lambda_p}{\sigma-\sigma_p}\,.
\end{equation*}
Recall that, from the form of $\lambda(\sigma_s)$ and on the support of the delta-functions, which will eventually be interpreted as the scattering equations, the limit 
\begin{equation*}
 \lim_{\sigma_s\rightarrow\sigma_a}\frac{\la \xi\,\lambda(\sigma_s)\ra}{\la s\,\lambda(\sigma_s)\ra}=\frac{\la\xi\,a\ra}{\la s\,a\ra}\,,\qquad a\in\{1,\dots,n\}\,.
\end{equation*}
Using the residue theorem and the support of the remaining scattering equations, the soft graviton Ward identity then takes the form
\begin{equation}
 \left\langle\cV_1\dots \cV_n \cV_s^0\right\rangle=\sum_{a=1}^n\frac{[as]\la \xi\,a\ra^2}{\la a\,s\ra\la \xi\, s\ra^2}\left\langle\cV_1\dots \cV_n\right\rangle\,,
\end{equation}
which can be identified straightforwardly as the soft graviton contribution. The soft graviton term thus arises from a specific charge generating supertranslation, which can be manifestly identified with the leading order expansion of an insertion of a soft graviton.\\

\subsubsection*{Subleading terms}

Expanding the integrated graviton vertex operator to first order in the soft momentum $s$ defines a superrotation,
\begin{equation}
\cV_s^1=\frac{1}{2\pi}\oint  \left( \frac{\la \xi \, \lambda(\sigma_s)\ra [\tilde\lambda(\sigma_s) \, \tilde \lambda_s]  [\mu(\sigma_s) \tilde\lambda_s]}{\la \xi \,  \lambda_s\ra \la \lambda_s\, \lambda (\sigma_s)\ra}+\frac{\la \xi\,\lambda(\sigma_s)\ra[\rho\,\tilde\lambda_s][\tilde\rho\,\tilde\lambda_s]}{\la\xi\,\lambda_s\ra\la\lambda_s\,\lambda(\sigma_s)\ra}\right) \,.
\end{equation}
Again, we can investigate the `Ward identity' associated to this superrotation, where we insert $Q_R$ in a correlation function of graviton vertex operators;
\begin{equation}
 \left\langle \tilde{\mathcal{V}}_{1}...\tilde{\mathcal{V}}_{k}\mathcal{V}_{k+1}...\mathcal{V}_{n}\cV_s^1\right\rangle\,.
\end{equation}
Using
\[
\left[\mu\left(\sigma_{s}\right)s\right]\bar{\delta}^{2}\left(\tilde{\lambda}_{i}-s_{i}\tilde{\lambda}\left(\sigma_{i}\right)\right)=\frac{1}{\sigma_{s}-\sigma_{i}}\tilde{\lambda}_{s}\cdot\frac{\partial}{\partial\tilde{\lambda}\left(\sigma_{i}\right)}\bar{\delta}^{2}\left(\tilde{\lambda}_{i}-s_{i}\tilde{\lambda}\left(\sigma_{i}\right)\right)+...
\]
we can calculate the correlation functions easily,
\begin{align*}
 &\left\langle \tilde{\mathcal{V}}_{1}...\tilde{\mathcal{V}}_{k}\mathcal{V}_{k+1}...\mathcal{V}_{n}\cV_s^1\right\rangle\\\
 &\qquad=\frac{1}{2\pi}\left\langle \oint \rd \sigma_s\sum_{i=1}^k \frac{\la \xi\,\lambda(\sigma_s)\ra [\tilde\lambda(\sigma_s)\,s]}{\la \xi\,s\ra \la \lambda(\sigma_s)\,s\ra}\frac1{\sigma_s-\sigma_i}\tilde\lambda_s\cdot\frac{\p}{\p\tilde\lambda(\sigma_i)} \tilde{\mathcal{V}}_{1}...\tilde{\mathcal{V}}_{k}\mathcal{V}_{k+1}...\mathcal{V}_{n}\right\rangle\\
 &\qquad\qquad + \frac{1}{2\pi}\Bigg\langle \oint \rd \sigma_s \sum_{p=k+1}^n \frac{\la \xi\,\lambda(\sigma_s)\ra }{\la \xi\,s\ra \la \lambda(\sigma_s)\,s\ra} I_R \;\tilde{\mathcal{V}}_{1}...\tilde{\mathcal{V}}_{k}\mathcal{V}_{k+1}...\widehat{\mathcal{V}}_{p}...\mathcal{V}_{n} \Bigg\rangle
\end{align*}
where the notation $\widehat{\mathcal{V}}_{p}$ indicates that the integrand of the corresponding vertex operator is omitted from the correlation function, still leaving the integration over the variable $\rd s_p/s_p^3$ and the scattering equation $\bd^2(\lambda_p-s_p\lambda(\sigma_p))$, and where
\begin{align*}
 I_R=[\tilde\lambda(\sigma_s)\,s]\frac{[s\,p]}{\sigma_s-\sigma_p} +(-1)^p\left([\tilde\rho(\sigma_p)\,p][\rho(\sigma_s)\,s]+[\tilde\rho(\sigma_s)\,s][\rho(\sigma_p)\,p]\right)\frac{[s\,p]}{\sigma_s-\sigma_p}.
\end{align*}
Trivially, the derivative $\tilde\lambda_s\cdot\frac{\p}{\p\tilde\lambda(\sigma_i)}$ can be taken to act on all vertex operators, as the only occurence of $\tilde\lambda(\sigma_i)$ is in the scattering equations. Note furthermore that the terms in $I_R$ can be obtained alternatively by acting with $\frac{[s\,p]}{\sigma_s-\sigma_p}\tilde\lambda_s\cdot\frac{\p}{\p\tilde\lambda_p}$ on the vertex operators in the correlation function, with the first term arising from the diagonal elements of $\HH_{pp}$, and the remaining terms from the off-diagonal contributions $\HH_{pq}$. We can thus rewrite the correlation function as 
\begin{align*}
  &\left\langle \tilde{\mathcal{V}}_{1}...\tilde{\mathcal{V}}_{k}\mathcal{V}_{k+1}...\mathcal{V}_{n}\cV_s^1\right\rangle\\\
 &\qquad=\frac{1}{2\pi}\left\langle \oint \rd \sigma_s\sum_{i=1}^k \frac{\la \xi\,\lambda(\sigma_s)\ra [\tilde\lambda(\sigma_s)\,s]}{\la \xi\,s\ra \la \lambda(\sigma_s)\,s\ra}\frac1{\sigma_s-\sigma_i}\tilde\lambda_s\cdot\frac{\p}{\p\tilde\lambda(\sigma_i)} \tilde{\mathcal{V}}_{1}...\tilde{\mathcal{V}}_{k}\mathcal{V}_{k+1}...\mathcal{V}_{n}\right\rangle\\
 &\qquad\qquad + \frac{1}{2\pi}\Bigg\langle \oint \rd \sigma_s \sum_{p=k+1}^n \frac{\la \xi\,\lambda(\sigma_s)\ra }{\la \xi\,s\ra \la \lambda(\sigma_s)\,s\ra} \frac{[s\,p]}{\sigma_s-\sigma_p}\tilde\lambda_s\cdot\frac{\p}{\p\tilde\lambda_p} \;\tilde{\mathcal{V}}_{1}...\tilde{\mathcal{V}}_{k}\mathcal{V}_{k+1}...\widehat{\mathcal{V}}_{p}...\mathcal{V}_{n} \Bigg\rangle
\end{align*}
Now the integral can be calculated straightforwardly, using the expicit expressions for $\lambda(\sigma)$ and $\tilde{\lambda}(\sigma)$, as well as the support of the delta-functions of the vertex operators. Thus the Ward identity for the superrotation charge obtained from the soft expansion of the graviton vertex operator gives the subleading terms of the soft limit,
\begin{equation}
 \left\langle \tilde{\mathcal{V}}_{1}...\tilde{\mathcal{V}}_{k}\mathcal{V}_{k+1}...\mathcal{V}_{n}\cV_s^1\right\rangle=\sum_{a=1}^n\frac{[a\,s]\la\xi\,a\ra}{\la a\,s\ra\la\xi\,s\ra}\tilde\lambda_s\cdot\frac{\p}{\p \tilde\lambda_a}\left\langle \tilde{\mathcal{V}}_{1}...\tilde{\mathcal{V}}_{k}\mathcal{V}_{k+1}...\mathcal{V}_{n}\right\rangle
\end{equation}

\subsubsection*{Sub-subleading terms}
Although there is no symmetry principle to protect the subsubleading terms in the soft expansion of the graviton vertex operators, we can still calculate the corresponding tree-level soft limits. In particular, the Ward identity for the diffeomorphism on ambitwistor space induced by the soft vertex operator to subsubleading order is given by
\begin{align*}
 &\left\langle \tilde{\mathcal{V}}_{1}...\tilde{\mathcal{V}}_{k}\mathcal{V}_{k+1}...\mathcal{V}_{n}\cV_s^{2}\right\rangle\\
 &\qquad= \frac1{2\pi i}\left\langle \oint \rd \sigma_s \frac{[\rho\,s][\tilde\rho\,s]}{\la s\,\lambda(\sigma_s)\ra}\;I_1\;\tilde{\mathcal{V}}_{1}...\tilde{\mathcal{V}}_{k}\mathcal{V}_{k+1}...\mathcal{V}_{n}\right\rangle\\
 &\qquad \qquad + \frac1{2\pi i}\left\langle \oint \rd \sigma_s \frac1{2}\frac{[\tilde\lambda(\sigma_s)\,s]}{\la s\,\lambda(\sigma_s)\ra}\;I_2\;\tilde{\mathcal{V}}_{1}...\tilde{\mathcal{V}}_{k}\mathcal{V}_{k+1}...\mathcal{V}_{n}\right\rangle
\end{align*}
where we have chosen to abbreviate the integrands by
\begin{align*}
 I_1&=\sum_{i=1}^k \frac1{\sigma_{si}}\tilde\lambda_s\cdot\frac{\p}{\p\tilde\lambda(\sigma_i)}+\sum_{p=k+1}^n \frac{[s\,p]}{\sigma_s-\sigma_p}\widehat\cV_p\\
 I_2&=\sum_{i,j=1}^k  \frac1{\sigma_{si}}\tilde\lambda_s\cdot\frac{\p}{\p\tilde\lambda(\sigma_i)} \frac1{\sigma_{sj}}\tilde\lambda_s\cdot\frac{\p}{\p\tilde\lambda(\sigma_j)} + \sum_{p,q=k+1; p\neq q}^n \frac{[s\,p]}{\sigma_{sp}}\frac{[s\,q]}{\sigma_{sq}}\widehat\cV_p\widehat\cV_q\\
 &\quad+ \sum_{p=k+1}^n\sum_{i=1}^k \frac{[s\,p]}{\sigma_{sp}}\frac1{\sigma_{si}}\tilde\lambda_s\cdot\frac{\p}{\p\tilde\lambda(\sigma_i)}\widehat\cV_p\,.
\end{align*}
Again, we have indicated by $\widehat\cV_p$ that the corresponding integrand of the vertex opertor is omitted from the correlation function. Calculating the residues and comparing the results to the derivatives obtained by acting with $\tilde{\lambda}_s\cdot\frac{\p}{\p\tilde\lambda_p}$ for $p\in\{k+1,\dots,n\}$, all unwanted residues cancel and the only contributions are coming from 
\begin{equation}
\begin{split}
 &\left\langle \tilde{\mathcal{V}}_{1}...\tilde{\mathcal{V}}_{k}\mathcal{V}_{k+1}...\mathcal{V}_{n}\cV_s^{2}\right\rangle\\
 &\qquad=\frac1{2}\sum_{a=1}^n \frac{[a\,s]}{\la a\,s\ra}\tilde{\lambda}_s\cdot\frac{\p}{\p\tilde\lambda_a}\tilde{\lambda}_s\cdot\frac{\p}{\p\tilde\lambda_a}\left\langle \tilde{\mathcal{V}}_{1}...\tilde{\mathcal{V}}_{k}\mathcal{V}_{k+1}...\mathcal{V}_{n}\right\rangle\,,
 \end{split}
\end{equation}
which is the subsubleading soft graviton contribution discovered in \cite{Cachazo:2014fwa}.

\bibliography{twistor-bib}  
\bibliographystyle{JHEP}

\end{document}